\documentclass[lettersize,journal]{IEEEtran}

\usepackage[T1]{fontenc}
\usepackage{cite}
\usepackage{amsmath,amssymb,amsfonts}
\usepackage{algorithmic}
\usepackage{algorithm}
\usepackage{graphicx}
\usepackage{textcomp}
\usepackage{xcolor}
\newcommand{\toolname}{N2NMatcher}
\def\BibTeX{{\rm B\kern-.05em{\sc i\kern-.025em b}\kern-.08em
		T\kern-.1667em\lower.7ex\hbox{E}\kern-.125emX}}

\usepackage{url}
\usepackage{subfigure}
\usepackage{multirow}
\usepackage{tabularx} 
\usepackage{enumitem}

\begin{document}
	
	\title{\toolname{}: Towards Inlining-Resilient Binary Decomposition and Module Matching}

	\author{Ang~Jia, He~Jiang, Zhipeng~Yang, Zhilei~Ren, Xiaochen~Li
		
		\thanks{Corresponding author: He~Jiang.}
		\thanks{This work has been submitted to the IEEE for possible publication. Copyright may be transferred without notice, after which this version may no longer be accessible.}
		
		\thanks{\textbullet~Ang~Jia, He~Jiang, Zhipeng~Yang, Zhilei~Ren, and Xiaochen~Li are with the School of Software, Dalian University of Technology, China. }

	}

	
	\maketitle
	
	\begin{abstract}
		Program-level Binary Code Similarity Analysis (BCSA) aims to identify semantically similar code regions across binary programs, serving as a fundamental technique for software plagiarism detection, vulnerability search, and malware analysis. Existing approaches often decompose binaries into modules following the structure of function call graphs (FCGs) and then match these modules by their contained functions. However, function inlining changes both FCG structures and binary function semantics, making existing decomposition and module matching methods less effective.
		
	In this work, we propose \toolname{}, an inlining-resilient framework for binary decomposition and module matching. We first conduct an empirical study to examine whether binaries still contain functions that provide stable module boundaries across compilation settings. \toolname{} learns to predict such functions as anchors using a hierarchical graph neural network that encodes binary ACFG--FCG representations built from opcode sequences, control-flow structures, and FCG calling context. It then performs anchor-bounded decomposition and matches the generated modules using learned module graph embeddings. Experimental results show that \toolname{} improves both the decomposition quality and module matching accuracy, compared to existing works, enabling more effective program-level BCSA.
		
	\end{abstract}
	

	\section{Introduction}
	
	\IEEEPARstart{M}{ost} software today is not developed entirely from scratch. Instead, developers rely on a range of open-source components to create their applications~\cite{kula2018developers}. According to a recent report~\cite{CodeReuse}, 97\% of software contains open-source components. Although code reuse helps to finish projects more quickly and reduce costs, improper reuse introduces security and legal risks~\cite{gkortzis2021software}. As reported in \cite{CodeReuse}, 65\% of organizations experienced a software supply chain attack in 2025. Moreover, downstream software often depends on closed-source binaries. Security and legal risks embedded in the binaries generated by upstream suppliers may be unintentionally transferred to downstream users.
	
	To address these code reuse issues, Program-level Binary Code Similarity Analysis (BCSA) identifies semantically similar code regions across different binary programs. In general, program-level BCSA consists of two stages: \textit{binary decomposition} and \textit{module matching}. In the binary decomposition stage, a binary is decomposed into modules that serve as the basic units for comparison, while module matching compares these modules across binaries to identify reused regions. Existing methods usually rely on function call graphs (FCGs) to decompose binaries and compare modules, assuming that reused code preserves similar FCG structures and function-level semantics.
	
\begin{figure*}[h]
		\centering
		\includegraphics[width=0.9\textwidth]{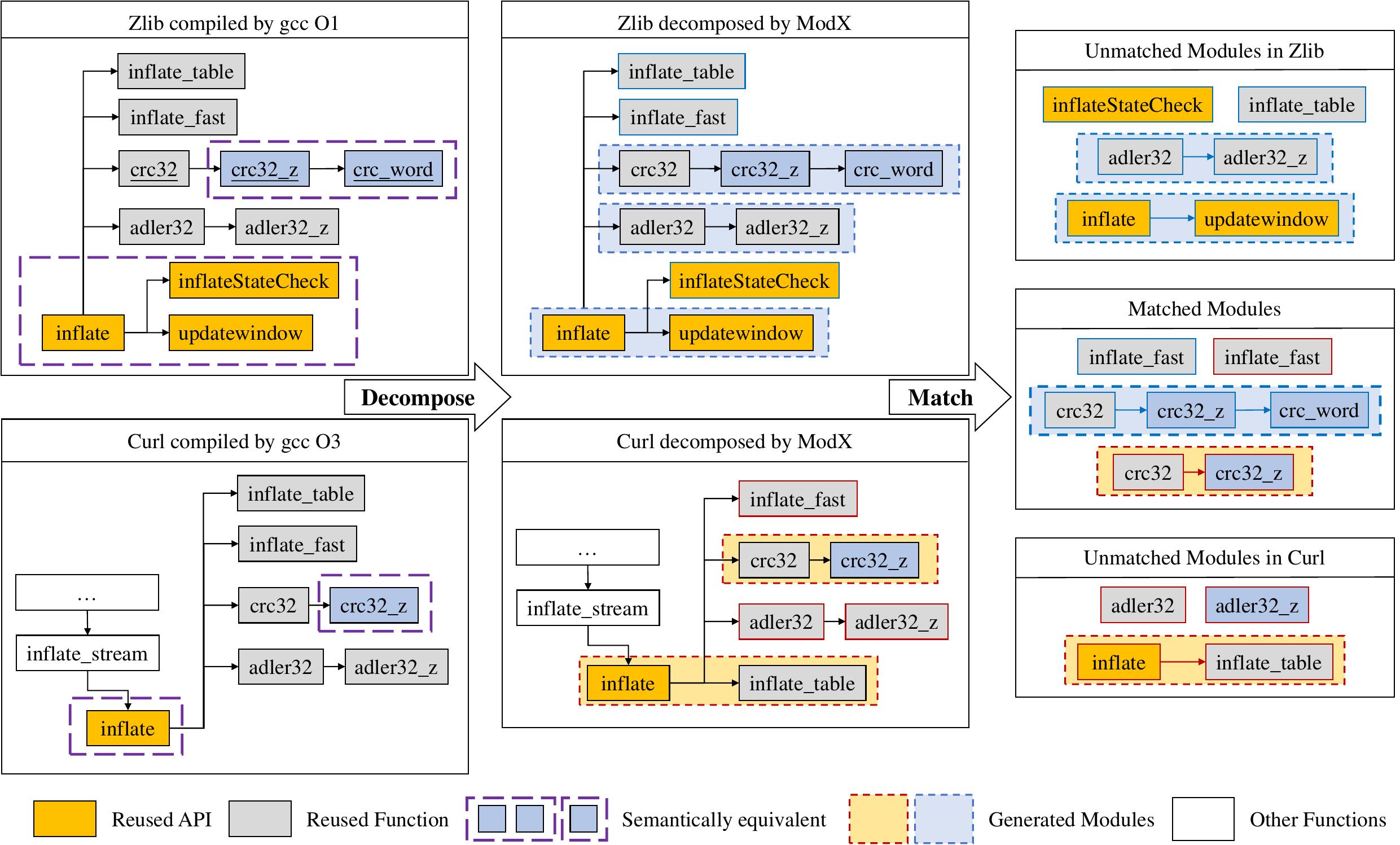}
		\vspace{-10pt}
		\caption{A motivating example for binary decomposition in program-level BCSA.}
		\label{fig:motivation_example}
	\end{figure*}

	However, as binaries are compiled using diverse compilers, optimizations, and architectures, their function call relationships undergo significant changes. Figure~\ref{fig:motivation_example} illustrates a reuse detection example between two software components: \textit{Zlib}~\cite{zlib} and \textit{Curl}~\cite{curl}. \textit{Zlib} is a widely reused compression library. \textit{Curl} integrates the \textit{inflate} module from \textit{Zlib} to handle compressed HTTP responses. In this example, the two components are compiled under different optimization settings: \textit{Zlib} is compiled with GCC at \textbf{O1}, while the \textit{Zlib} code integrated into \textit{Curl} is compiled with GCC at the more aggressive \textbf{O3} level. This difference in optimization leads to \textbf{distinct inlining decisions}.
	
	As shown in Figure~\ref{fig:motivation_example}, function inlining changes both the \textbf{topology of FCGs} and the \textbf{semantics of their nodes}. For example, the function \textit{crc32\_z} in \textit{Curl} inlines \textit{crc\_word} from \textit{Zlib}, so \textit{crc\_word} no longer exists as a standalone binary function in \textit{Curl}. Similarly, \textit{inflate} inlines \textit{inflateStateCheck} and \textit{updatewindow}; as a result, a single binary function in \textit{Curl} may correspond to multiple binary functions in \textit{Zlib}. This variance directly undermines the core assumption of existing binary decomposition methods. Once the decomposed modules are no longer semantically aligned, downstream module matching also becomes unreliable.
	
	When applying existing state-of-the-art decomposition methods, such as ModX~\cite{yang2022modx}, to this example, we observe that they face two main challenges. \textbf{Challenge 1: unstable module boundaries.} Function inlining changes the FCG structures used for modularization, so modules generated from the two binaries may no longer cover the same source-level regions. \textbf{Challenge 2: unstable function-level semantics.} ModX also relies on function-level semantics to calculate module similarity, but inlining changes semantics in the binary functions. For example, \textit{crc32\_z} in \textit{Curl} contains the semantics of both \textit{crc32\_z} and \textit{crc\_word} in \textit{Zlib}. Although these modules cover related source semantics, ModX may fail to identify their equivalence.
	
	These mismatches show that existing methods cannot effectively handle the module boundary changes and function semantic changes caused by function inlining. To resolve Challenge 1, we first conduct an empirical study to determine whether binaries still contain functions that can serve as stable module boundaries across compilation settings. The study shows that although function-level equivalence is often disrupted, source-verified stable boundary functions still exist and can delimit source-equivalent binary regions.
	
	Motivated by this observation, we propose \textit{\toolname{}}, an inlining-resilient framework for binary decomposition and module matching. The key insight behind \toolname{} is to identify functions that can serve as stable boundaries and use them as anchors. To identify anchors in a binary, \toolname{} first constructs a binary ACFG--FCG representation by combining function-level attributed control-flow graphs (ACFGs) with interprocedural call relationships. It then applies a hierarchical graph neural network to encode opcode, control-flow, and calling-context information and predict which FCG nodes are anchors. Following the predicted anchors, \toolname{} decomposes a binary into anchor-bounded regions whose boundaries are less sensitive to inlining-induced FCG changes.
	
	To address Challenge 2, \toolname{} further performs matching at the module level. Instead of relying only on unstable one-to-one function-level similarities, \toolname{} represents each generated module as a graph and compares the resulting module graph embeddings. This design reduces the impact of inlining-induced function granularity changes and enables program-level BCSA to compare module similarities rather than isolated functions.
	

	Our main contributions are listed as follows:
	
	\begin{enumerate}
		\item To the best of our knowledge, we propose the first inlining-resilient framework that supports binary decomposition and module matching for program-level BCSA.
		
		\item We conduct an empirical study to reveal that source-verified stable boundary nodes widely exist under compilation variance and can serve as effective boundaries for binary decomposition.

		\item Building on this observation, we propose \textit{\toolname{}}, which predicts anchors with a hierarchical ACFG--FCG graph neural network, decomposes binaries through anchor-bounded traversal, and performs module matching using module graph embeddings.

		\item We evaluate \textit{\toolname{}} under project-level cross-validation. The results show \textit{\toolname{}} improves both the decomposition quality and module matching accuracy when compared to existing works.
		
	\end{enumerate}
	
	To facilitate further research, we have made the implementation of \toolname{} publicly available on GitHub.\footnote{\url{https://github.com/island255/N2NMatcher}}

	
	\section{Related Work}
	
	Our work studies program-level BCSA under function inlining. In this setting, binary analysis often involves two stages: decomposing a binary into module-level regions and comparing these regions across binaries. Function inlining affects both stages because it changes the topology of FCGs used for decomposition and the semantic granularity of binary functions used for matching. Therefore, we discuss related work in two areas: binary decomposition for program-level BCSA and binary code similarity analysis under function inlining.
	
	\subsection{Binary Decomposition for Program-level BCSA}
	
	Binary decomposition is an important step in program-level binary analysis, especially for C/C++ third-party library detection and component reuse detection~\cite{li2023libam, ISRD, dong2024libvdiff, tang2022libdb, zhu2022bbdetector, yang2022modx, sun2023moddiff, karande2018bcd, guo2023searching, tang2020libdx}. Compared with Java library detection~\cite{backes2016reliable, li2017libd, ma2016libradar, zhan2021atvhunter, zhan2020automated, zhang2019libid, zhang2018detecting}, C/C++ binary decomposition faces more diverse compilation settings and architectures, making module boundaries harder to preserve.

	Existing fine-grained binary decomposition methods can be broadly divided into two classes: anchor-based methods and clustering-based methods.
	
	
	Anchor-based methods~\cite{li2023libam, ISRD, dong2024libvdiff, tang2022libdb} first identify highly similar function pairs as anchor nodes, and then follow the function call relationships around these anchors to search for additional matched functions. ISRD~\cite{ISRD} identifies functions with identical instructions or identical library calls as anchors, and then compares functions that share many anchor neighbors. LibDB~\cite{tang2022libdb} selects similar function pairs whose similarity is larger than 80\%, and then uses the FCG structure to examine the matched functions. LibAM~\cite{li2023libam} first identifies highly similar function pairs as anchors, extends these anchors to construct function areas, and finally compares function areas to identify reused third-party libraries (TPLs). Though anchor-based methods use different strategies to detect reused TPLs, they all rely on the local structure around anchor nodes to match more functions.

	
	Clustering-based methods~\cite{yang2022modx, sun2023moddiff, karande2018bcd, guo2023searching} first cluster binary functions into modules, and then compare these modules to identify reused TPLs. BCD~\cite{karande2018bcd} uses three properties, including code locality, data references, and function calls, to construct a graph for clustering. Then BCD uses Newman's generalized community detection algorithm~\cite{newman2004fast} to group binary functions into modules.
	ModX~\cite{yang2022modx} first defines a module quality score to assess the coherence of the function clusters, and then it starts to group individual functions to form modules while maximizing the overall module quality score. 
	BMVul~\cite{guo2023searching} introduces directed binary modularization (DBM), an overlapping community detection algorithm for binary function clustering. 
	Though clustering-based methods use different features to construct their clustering graphs, the FCG is usually the core structure from which clustering starts.

	\subsection{BCSA Under Function Inlining}
	

	Existing works~\cite{bingo, asm2vec, jia20231, jia2022comparing, jia2024cross, lin2024reifunc, sha2025optrans, qiu2015library, qiu2015using, ahmed2021learning, binosi2023bino, lin2023fsmell, dall2022highliner} have taken preliminary steps toward binary code analysis under function inlining. The first binary function similarity detection work that considers function inlining is Bingo~\cite{bingo}, which summarizes several patterns that a caller function may use to inline its callee functions and conducts inlining to simulate the resulting binaries. Asm2Vec~\cite{asm2vec} also adopts Bingo's strategies to handle function inlining. However, their simulation-based inlining strategies have been proven inaccurate~\cite{jia20231} and thus cannot fully address the challenges that function inlining brings.
	
	Jia et al.~\cite{jia20231} conducted the first systematic empirical study to investigate the effect of function inlining on binary function similarity analysis. They evaluated four existing works on an inlining-aware dataset and observed a 30\%--40\% performance loss when detecting functions affected by inlining. To tackle these challenges, they proposed two methods, O2NMatcher~\cite{jia2022comparing} and CI-Detector~\cite{jia2024cross}, for binary-to-source and binary-to-binary function-level similarity detection, respectively.
	
	Abusabha et al.~\cite{abusabha2025deep} systematically analyzed how inlining distorts static features in machine-learning (ML)-based binary analysis and showed that compiler settings can be exploited to craft evasive binary variants. Moreover, several works aim to identify inlined functions. ReIFunc~\cite{lin2024reifunc} identifies inlined functions by detecting repeated code fragments in binaries. A number of works~\cite{qiu2015library, qiu2015using, ahmed2021learning, binosi2023bino} focus on identifying inlined library functions. Besides, OpTrans~\cite{sha2025optrans} uses a binary rewriting technique to reoptimize binaries toward similar inlining decisions.
	
	However, the above works mostly focus on function-level binary similarity analysis. Jia et al.~\cite{jia2025towards} conducted an empirical study on binary decomposition under compilation variance and found that existing decomposition methods suffer from under-aggregation and over-aggregation failures. In this work, we propose a binary decomposition method designed to be robust to function inlining. Our work provides a new perspective for subsequent research from a program-level view.

	
	\section{Empirical Study}
	\label{sec:rq1}
	\label{sec:stable_node_study}
	
	The motivating example reveals that function inlining introduces changes at two levels: it changes the topology of FCGs used for decomposition and alters the semantic granularity of binary functions used for function-level matching. As a result, modules generated from different compilation variants may no longer cover the same source-level semantics, and individual binary functions may no longer preserve one-to-one semantic equivalence. In this section, we conduct an empirical study to investigate whether stable semantic regions still exist under such changes and, if so, at what binary granularity semantic equivalence can be preserved. We therefore formulate the following study question:
	
	\begin{quote}
		\textbf{How does function inlining affect semantic equivalence across compilation variants, and can semantic equivalence be preserved at a certain binary granularity?}
	\end{quote}
	
	We decompose this question into three study questions that progressively examine FCG equivalence, function-level semantic equivalence, and region-level semantic equivalence:
	
	\begin{itemize}
		\item \textbf{SQ1.} To what extent do FCGs remain structurally equivalent across compilation variants?
		\item \textbf{SQ2.} How does function inlining affect function-level semantic equivalence?
		\item \textbf{SQ3.} Can semantic equivalence be preserved at a coarser binary-region granularity?
	\end{itemize}

	\subsection{Study Design}
	
	\subsubsection{Dataset}
	
	We conduct the empirical study on x86-64 binaries from the link-time optimization (LTO) subset of BinKit~\cite{Binkit,BinkitGithub}. We choose this dataset for two reasons. First, BinKit is a widely used BCSA benchmark and has also been adopted in recent binary decomposition evaluation~\cite{jia2025towards}, making it suitable for studying decomposition under controlled compilation variance. Unlike datasets such as BinaryCorp~\cite{10.1145/3533767.3534367}, which mainly emphasize scale and diversity for function-level similarity learning, BinKit provides an explicit LTO subset for studying decomposition under aggressive whole-program optimization. 
	Second, the LTO subset is closely aligned with our research objective because it exposes substantial FCG changes. Link-time optimization enables interprocedural optimization across compilation units and supports whole-program optimization, including cross-file inlining~\cite{GCC_inline}. Since prior work has shown that function inlining significantly affects binary similarity analysis~\cite{jia20231}, this subset allows us to further investigate how inlining affects program-level binary decomposition and module matching.

	The dataset contains 84 executable targets derived from 28 projects. Each executable target is compiled using 9 compiler versions, including Clang 4.0, 5.0, 6.0, 7.0, and GCC 4.9.4, 5.5.0, 6.4.0, 7.3.0, 8.2.0, with four optimization levels, including O0, O1, O2, and O3. This process produces 3,024 binary variants in total. All binaries are compiled with debug information to facilitate labeling. In this empirical study, we use IDA Pro~\cite{IDAPro} to disassemble these binaries and construct one FCG for each binary. In summary, we obtain 3,024 FCGs with 1,403,680 nodes and 7,417,670 edges.

	\subsubsection{Binary-to-Source Function Mapping}
	
	We construct binary-to-source function mappings from binaries compiled with debug information to provide a source-level semantic oracle for the empirical study. Specifically, we extract records that map binary instruction addresses to source lines from debug-line tables. We also extract source function ranges from the corresponding source files. A binary function is mapped to a source function if its instructions map to source lines within the source function range. For a binary function $b$, let $M(b)$ denote the set of source functions mapped to $b$. If $|M(b)|>1$, we treat $b$ as an inlining-affected binary function because its recovered binary body contains source semantics from multiple source functions. 
	
	\subsubsection{Binary-to-Binary Function Mapping}
	
	Based on the source sets $M(b)$ obtained above, we construct binary-to-binary function mappings for binaries compiled from the same project target. Given two binary functions $b_i$ and $b_j$ from two compilation variants, we count them as an exact function-level match under our source-set oracle when $M(b_i)=M(b_j)\neq\emptyset$. When their source sets overlap but are not equal, the corresponding semantics are only partially aligned at the function level and may require region-level aggregation. For a binary region $R$, we define $M(R)=\bigcup_{b\in R}M(b)$ and use this source set to analyze the semantic coverage of binary regions.
	
	Unless otherwise stated, the function-level metrics in this section are computed over binary functions covered by the binary-to-source mapping. This restriction ensures that the reported results are supported by source-level semantic evidence. Binary FCG structural statistics, such as the numbers of recovered nodes and edges, are computed over the whole recovered FCGs.
	
	\subsection{SQ1: Impact on FCG Structures}
	
\begin{figure}[t]
		\centering
		\subfigure[FCG nodes]{
			\includegraphics[width=0.96\columnwidth]{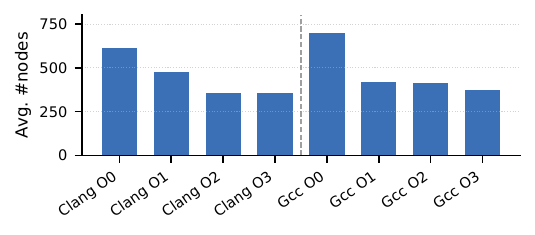}
		}\\[-0.5ex]
		\subfigure[FCG edges]{
			\includegraphics[width=0.96\columnwidth]{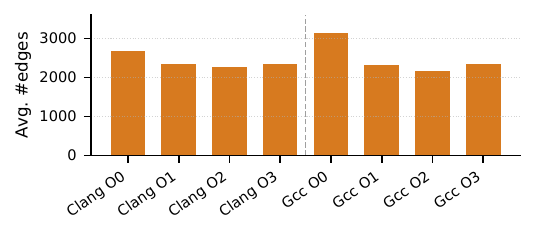}
		}
		\caption{FCG structural changes across compiler-optimization settings.}
		\label{fig:config_fcg_summary}
	\end{figure}

	We first examine whether recovered FCGs remain structurally equivalent across optimization variants. Function inlining is one mechanism that can remove standalone callees and rewire call relationships through their callers, thereby affecting both the number of visible FCG nodes and the number of call edges. Figure~\ref{fig:config_fcg_summary} reports these two structural statistics under eight compiler-optimization settings. For each compiler-optimization setting, we aggregate the results across different versions of the same compiler family.
	
	As shown in Figure~\ref{fig:config_fcg_summary}(a), the average number of FCG nodes decreases consistently from O0 to optimized settings. Clang decreases from 614.06 nodes at O0 to about 356 nodes at O2/O3, while GCC decreases from 697.43 nodes at O0 to 372.33 nodes at O3. This confirms that many source-level functions no longer remain as standalone binary functions after optimization. Figure~\ref{fig:config_fcg_summary}(b) shows that FCG edges also change substantially, although their trend is less monotonic than that of FCG nodes. For example, GCC edges decrease from 3,130.91 at O0 to 2,152.17 at O2, but rise to 2,348.68 at O3. This suggests that inlining not only removes nodes but also reshapes call relationships among the remaining binary functions.

	We further compare all six unordered optimization-level pairs for every executable target and exact compiler version, yielding 4,536 FCG comparisons. Isomorphic directed graphs must have equal numbers of nodes and edges. In 4,429 comparisons (97.64\%), at least one of these counts differs, which is sufficient to prove structural non-equivalence. All 2,268 comparisons between O0 and an optimized variant differ in at least one count. The remaining 107 comparisons have equal node and edge counts, but count equality is necessary rather than sufficient for graph isomorphism; we therefore leave these cases undecided instead of treating them as equivalent.

	\subsection{SQ2: Impact on Binary Functions}
	
\begin{figure}[t]
		\centering
		\includegraphics[width=0.98\columnwidth]{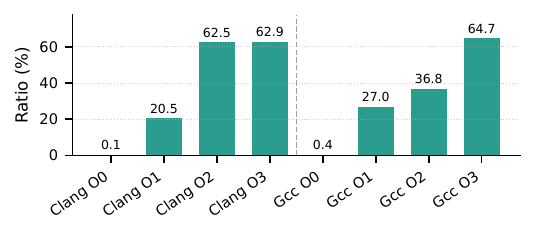}
		\caption{Inlining-affected binary functions across compiler-optimization settings.}
		\label{fig:inline_affected_ratio}
	\end{figure}

	We next analyze how function inlining affects binary functions themselves. We first measure how many recovered binary functions are affected by inlining. Following the binary-to-source mapping definition, a binary function is treated as inlining-affected when it maps to multiple source functions. Figure~\ref{fig:inline_affected_ratio} reports the ratio under eight compiler-optimization settings. The ratio is very low at O0, with 0.08\% for Clang and 0.35\% for GCC. It then increases sharply as optimization becomes stronger. For Clang, the ratio reaches 20.47\% at O1 and more than 62\% at O2/O3. For GCC, it increases from 26.96\% at O1 to 64.71\% at O3. These results show that optimized binaries contain many binary functions whose semantics are produced by merging multiple source functions.
	
	The above results show that many optimized binary functions are produced by merging multiple source functions. Since function inlining changes the semantic content of individual binary functions, methods whose module matching relies on function-level similarities, such as ModX, can also be affected in program-level BCSA. Therefore, we construct binary pairs under three common cross-compilation scenarios, ordered from easier to harder: cross-compiler same-optimization, same-compiler cross-optimization, and cross-compiler cross-optimization. Under these scenarios, we verify whether function-level semantic equivalence is still preserved. 
	
	For each pair of binaries from the same project target, if a query binary function has the same source set as a function in the other binary, we count it as an exact function-level equivalent. If no exact equivalent exists but the source set overlaps with at least one function in the other binary, we count it as partial function-level overlap. Such partial overlap indicates that the corresponding source semantics are still present in the other binary, but they are split, merged, or absorbed differently at the binary-function level. These cases indicate where region-level aggregation may recover semantic alignment beyond individual binary functions.
	
\begin{figure}[t]
		\centering
		\includegraphics[width=0.98\columnwidth]{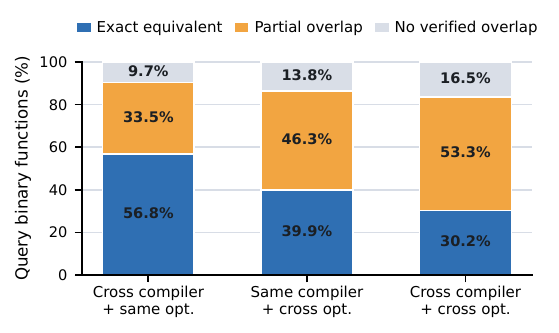}
		\caption{Function-level alignment across compilation variants.}
		\label{fig:function_alignment}
	\end{figure}

	Figure~\ref{fig:function_alignment} reports the results from 1,000 sampled binary pairs for each cross-compilation setting in the LTO dataset, with every pair evaluated in both directions. A large fraction of query functions no longer have exact one-to-one counterparts but still partially overlap with functions in the other binary: 33.5\% in cross-compiler same-optimization pairs, 46.3\% in same-compiler cross-optimization pairs, and 53.3\% in cross-compiler cross-optimization pairs. Meanwhile, the ratio of exact function-level equivalents drops from 56.8\% under cross-compiler same-optimization to 39.9\% under same-compiler cross-optimization and 30.2\% under cross-compiler cross-optimization. These cases indicate that equivalent source semantics are still present, but they are increasingly split, merged, or absorbed across binary functions rather than preserved as one-to-one binary functions.
	
	The remaining category, \textit{No verified overlap}, does not imply that the source semantics are absent from the target binary. It means that, under our debug-line-based oracle, none of the source functions mapped to the query binary function can be verified in any target-side binary function. For example, in \texttt{plotutils-2.6/ode.elf}, the GCC~8.2.0 O0 binary contains a recovered function \texttt{intpr} mapped to \texttt{ode/stperr.c:intpr} at lines 100--116. In the Clang~5.0 O2 binary of the same target, no recovered binary function maps to \texttt{intpr}, although the same source file still has verified mappings for other functions such as \texttt{maxerr}, \texttt{lowerror}, and \texttt{hierror}. This case is therefore counted as \textit{No verified overlap}. Such cases can arise when optimization eliminates, absorbs, or rewrites small functions without preserving debug-line information that can be attributed back to the original source function.
	
	\subsection{SQ3: Region-Level Semantic Equivalence}
	
	In SQ3, we investigate whether semantic equivalence can be preserved at a coarser binary granularity. For a binary region $R$, we define its source region as the union of source functions mapped from its binary functions, i.e., $S(R)=\bigcup_{b\in R}M(b)$. Two binary regions are source-equivalent if they map to the same source region. Under this view, binary decomposition should recover binary regions whose source regions remain stable across compilation variants.
	
	Such source-equivalent regions require stable boundary functions. In the motivating example, some source functions are absorbed by inlining, such as \textit{crc\_word}, \textit{inflateStateCheck}, and \textit{updatewindow}; these functions cannot mark stable region boundaries because they disappear as standalone binary functions in the optimized \textit{Curl} binary. In contrast, caller-side functions such as \textit{crc32\_z} and \textit{inflate} remain as the origin source functions of recovered binary functions and can delimit regions that contain their inlined callees. This suggests two observable properties of stable boundary functions: first, at least one observed compilation configuration recovers them as caller-side binary functions, possibly containing inlined callees; second, no observed configuration places them as inlined callees inside binary functions rooted in other source functions.
	
	We formalize these properties using binary-to-source mappings. For a binary function $b$, let $M(b)$ be its mapped source-function set and let $o(b)$ be its \textit{origin source function} (OSF), i.e., the source function that contributes the main semantics of the standalone binary function. If $b$ maps to one source function, the source function is treated as the OSF. If $b$ maps to multiple source functions, we treat the case as an inlining case: the source function whose name matches the binary function name is treated as the caller-side OSF, while the other mapped source functions are treated as inlined callee-side semantics. For a group $g$ of compilation variants generated from the same project target, let $C_g$ denote its compilation configurations. For each configuration $c \in C_g$, let $B_c$ denote its recovered binary functions. For a source function $s$, we define $B^{\mathrm{root}}_c(s)$ as the binary functions in $B_c$ whose OSF is $s$, and $B^{\mathrm{inl}}_c(s)$ as the binary functions in $B_c$ where $s$ is not the OSF but is inlined into the function whose OSF is $o(b)$:
	
	\begin{equation}
		B^{\mathrm{root}}_c(s)=\{b \in B_c \mid o(b)=s\},
	\end{equation}
	\begin{equation}
		B^{\mathrm{inl}}_c(s)=\{b \in B_c \mid s \in M(b) \land o(b)\neq s\}.
	\end{equation}
	
	\noindent\textbf{Definition 1 (Source-verified stable boundary node).}\makeatletter\def\@currentlabel{1}\label{def:stable_boundary_node}\makeatother
	A source function $s$ defines source-verified stable binary boundary nodes in a group $g$ if it satisfies the following condition:
	\begin{equation}
		\left(\exists c\in C_g:\ |B^{\mathrm{root}}_c(s)|\geq 1\right)
		\quad\land\quad
		\left(\forall c\in C_g:\ B^{\mathrm{inl}}_c(s)=\emptyset\right).
	\end{equation}
	The first term requires $s$ to appear as the OSF of at least one recovered binary function in the compilation group. The second term requires that, across all observed configurations, $s$ is never inlined into a binary function whose OSF is another source function. 
	
	Stability therefore refers to the boundary role of $s$ whenever it is recovered rather than requiring a binary instance in every configuration.	
	For each source function $s$ satisfying Definition~\ref{def:stable_boundary_node} and each configuration $c\in C_g$, every binary function in $B^{\mathrm{root}}_c(s)$ is labeled as a stable boundary node in the binary FCG. Thus, the label requires $s$ to occur as the OSF of the node and never as an inlined callee under another OSF in any observed configuration. The labeled binary function may nevertheless contain other source functions as inlined callees.

\begin{figure}[t]
		\centering
		\includegraphics[width=0.98\columnwidth]{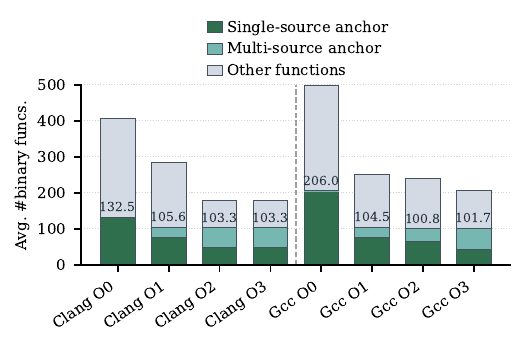}
		\caption{Composition of stable boundary nodes in the LTO dataset.}
		\label{fig:stable_boundary_composition}
	\end{figure}

	Although function inlining creates many inlining-affected binary functions, source-verified stable boundary nodes are still widely present. Among 855,226 FCG nodes with debug-line source labels, 364,896 nodes are source-verified stable boundary nodes, accounting for 42.67\% of labeled nodes and 26.00\% of all FCG nodes. Figure~\ref{fig:stable_boundary_composition} further shows that stable boundary nodes persist across compiler-optimization settings, while other functions shrink more sharply under optimization. Its eight compiler-optimization bars report average binary-function counts per binary. We split stable boundary nodes into single-source and multi-source cases: a single-source stable node maps to only its OSF, while a multi-source stable node maps to its OSF plus one or more inlined callees; the remaining source-labeled binary functions are grouped as other functions. Among all stable boundary nodes, 267,365 are single-source and 97,531 are multi-source, meaning that 26.73\% of stable boundaries contain inlined semantics while still preserving a stable origin.
	
	We further verify whether these stable boundaries actually delimit equivalent regions. Using the same 3,000 sampled binary pairs as SQ2, evaluated in both directions and yielding 6,000 directional evaluations, we construct regions delimited by source-verified stable boundary nodes and compare each query region with the most similar region in the other binary. Because optimization may eliminate source functions, the comparison uses only source functions observed in both binaries. For a pair $(B_q,B_t)$, let $U_{q,t}$ be their common mapped source-function set, and let $S_{q,t}(r)=S(r)\cap U_{q,t}$. Regions with $S_{q,t}(r)=\emptyset$ are excluded. Given a query region $r_q$ and a target region $r_t$, we define the similarity as:
	\begin{equation}
		\mathrm{Overlap}_{\mathrm{sf}}(r_q,r_t)=
		\frac{2|S_{q,t}(r_q)\cap S_{q,t}(r_t)|}
		{|S_{q,t}(r_q)|+|S_{q,t}(r_t)|}.
	\end{equation}
	This source-level overlap score ranges from 0 to 1, where 1 means that the two regions cover the same verified source-function set and 0 means that they share no verified source functions. For each query region, we retain the highest score over all target regions. Across 503,786 source-relevant query regions, the mean of these per-query scores is 0.938; 71.76\% of regions have an exact counterpart, and 89.99\% reach a score of at least 0.8. 
	Even in the hardest cross-compiler cross-optimization setting, the mean score remains 0.923, with 87.51\% of regions reaching 0.8. These results indicate that semantic equivalence can largely be preserved at the coarser granularity of binary regions obtained through stable-boundary-guided decomposition, even when individual binary functions no longer align one-to-one across compilation variants.
	
	\subsection{Answer to SQ1--SQ3}
	
	Overall, function inlining disrupts function-level semantic alignment, but it does not eliminate semantic equivalence entirely. First, recovered FCGs are generally structurally non-equivalent across optimization variants, so binary decomposition should not assume stable FCG topology. Second, function inlining changes binary function granularity, making many cross-compilation function pairs only partially overlap at the source level. Third, semantic equivalence can still be preserved at a source-consistent region granularity, with 42.67\% of labeled FCG nodes serving as stable boundaries under our definition. These findings motivate learning stable boundary nodes and using them to guide binary decomposition, as described in the next section. The resulting source-consistent modules provide a more reliable basis for downstream module matching.
	
	\section{\toolname{}}
	\label{sec:method}
	
	\begin{figure*}[t]
		\centering
		\vspace{-5pt}
		\includegraphics[width=0.78\textwidth]{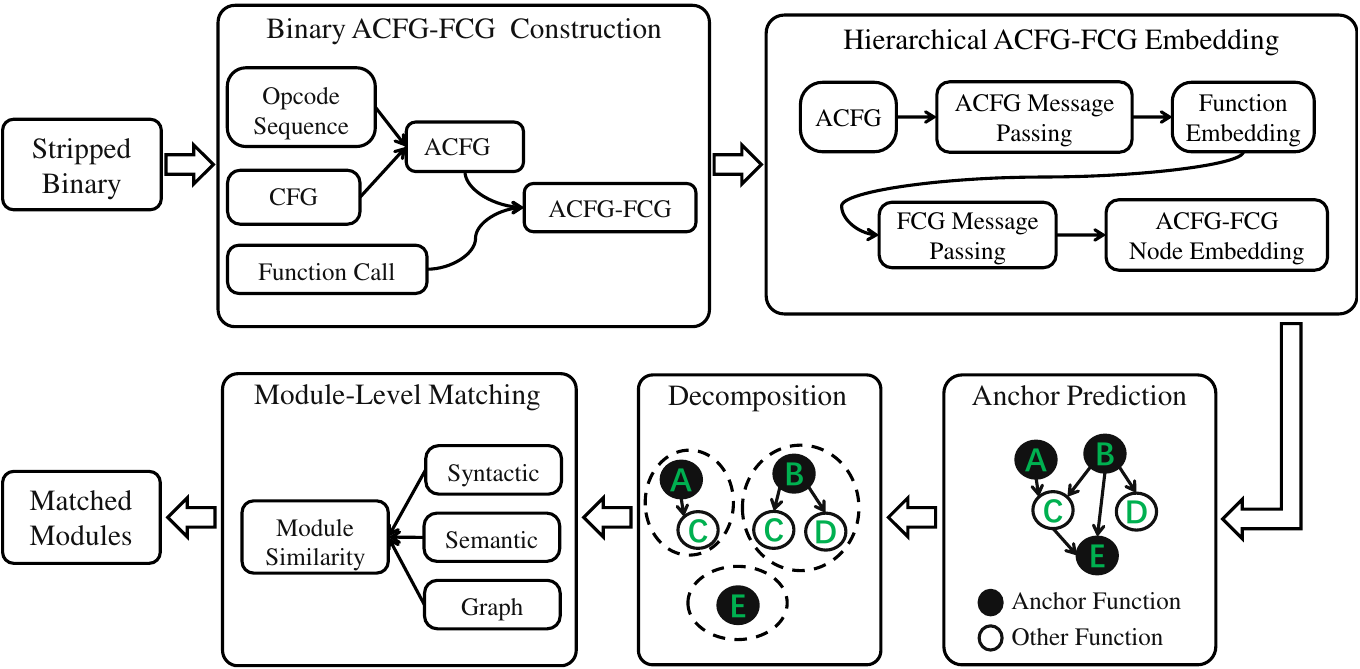}
		\vspace{-8pt}
		\caption{Overview of \toolname{} for binary decomposition and module matching.}
		\label{fig:overview}
		\vspace{-8pt}
	\end{figure*}
	
	This section presents \toolname{}, an inlining-resilient framework for binary decomposition and module matching. The key idea is to predict binary FCG nodes that can serve as stable boundaries. We refer to these predicted nodes as anchors and use them to guide decomposition. Instead of clustering the whole FCG according to global topology, \toolname{} predicts anchors from binary graph representations and then constructs anchor-bounded modules.
	
	Figure~\ref{fig:overview} presents the overall workflow of \toolname{}. Given a stripped binary, \toolname{} first constructs an ACFG--FCG representation by combining function-level attributed control-flow graphs with interprocedural call relationships. It then embeds this representation with a hierarchical ACFG--FCG encoder, which produces contextual representations for FCG nodes. Based on these embeddings, \toolname{} predicts anchors, decomposes the FCG into anchor-bounded modules, and compares the generated modules by integrating Syntactic Similarity, Semantic Similarity, and Graph Similarity. Source code and debug information are used only for constructing training labels and evaluation oracles. During deployment, \toolname{} requires only a stripped binary as input.

	\subsection{Binary ACFG--FCG Construction}
	
	\textbf{Instruction Normalization.}
	Given a stripped binary, \toolname{} first converts disassembly into a compact instruction representation. For each instruction, we keep only its opcode mnemonic in lowercase as the normalized token and discard operand-level details such as registers, concrete addresses, and immediate values. This normalization reduces sensitivity to register allocation, address layout, and compiler-specific instruction changes, while retaining a compact record of the operations observed in each basic block. 
	
	\textbf{Basic Block Representation.}
	For each basic block, the normalized opcode mnemonics are mapped into a deterministic hash-based token space and then converted into learnable embeddings. Mean pooling produces an order-invariant summary of the opcode content, which is projected to initialize the basic-block node representation. The resulting representations are then updated through ACFG-level message passing.
	
	\textbf{ACFG Construction.}
	For each recovered binary function $v$, \toolname{} builds an attributed control-flow graph (ACFG) $G_v=(V_v,E_v)$. Each node in $V_v$ corresponds to a basic block and is attributed with the normalized opcode sequence extracted from that block. Each directed edge $(b_i,b_j)\in E_v$ denotes an intraprocedural control-flow transfer from basic block $b_i$ to $b_j$. The hierarchical encoder uses the normalized opcode sequences as ACFG node attributes. Thus, the ACFG captures both local instruction behavior and intraprocedural control-flow structure.
	
	\textbf{FCG Construction.}
	At the interprocedural level, \toolname{} constructs a function call graph (FCG) $G_f=(V_f,E_f)$ for the whole binary. Each node $v\in V_f$ corresponds to a recovered binary function. The directed edges in $E_f$ come from two sources: direct call instructions and direct cross-function jumps. For a direct call inside function $u$, \toolname{} adds an edge $(u,v)$ when the call target resolves to an address belonging to function $v$. For a direct cross-function jump inside function $u$, \toolname{} adds a tail-call edge $(u,v)$ when the statically resolved jump target belongs to a different recovered function $v$. This tail-call-aware FCG captures both ordinary calls and tail-call-style interprocedural transfers introduced by compiler optimizations.
	
	\textbf{ACFG--FCG Construction.}
	\toolname{} then attaches each function-level ACFG $G_v$ to its corresponding FCG node $v$. The resulting ACFG--FCG is a nested binary representation: the FCG captures interprocedural caller--callee relationships, while each attached ACFG captures the opcode content and intraprocedural control-flow structure of a recovered function. This nested representation provides the hierarchical encoder with both function-body structure and calling-context information, which are important for identifying stable module boundaries.

	\subsection{Hierarchical ACFG--FCG Embedding}
	
	The binary ACFG--FCG is a nested graph: each FCG node corresponds to a recovered function and is associated with an ACFG describing its intraprocedural structure. \toolname{} therefore uses a hierarchical encoder with two message-passing levels. The lower level propagates information within each ACFG to derive a function embedding for each recovered function. The upper level uses these function embeddings as initial FCG node representations and propagates them over FCG edges to incorporate calling-context information.
	
	At the ACFG level, normalized opcodes are mapped into a deterministic hash-based token space, and their embeddings are learned during model training. For a nonempty basic block $b$ with opcode sequence $(o_1,o_2,\ldots,o_m)$, \toolname{} looks up opcode embeddings, applies mean pooling over the tokens, and projects the pooled vector into the initial block representation $h_b^{(0)}$:
	\begin{equation}
		h_b^{(0)}=\mathrm{ReLU}\left(W_o\cdot \frac{1}{m}\sum_{i=1}^{m}\mathrm{Emb}(o_i)\right).
	\end{equation}
	The function-body encoder then applies message passing over the ACFG. For each basic block, it aggregates representations from adjacent basic blocks and combines them with the block's own state:
	\begin{equation}
		h_b^{(\ell+1)}=\mathrm{ReLU}\left(W_s h_b^{(\ell)}+
		W_n\cdot \mathrm{Mean}_{u\in \mathcal{N}_{A}(b)} h_u^{(\ell)}\right),
	\end{equation}
	where $\mathcal{N}_{A}(b)$ denotes the basic blocks adjacent to $b$ in the symmetrized message-passing graph. After two ACFG message-passing layers, the function representation is obtained by mean pooling over all basic blocks in the function:
	\begin{equation}
		z_v=\mathrm{Mean}_{b\in V_v} h_b^{(L_A)}.
	\end{equation}
	The vector $z_v$ is the ACFG embedding of function $v$. It captures both local opcode patterns and intraprocedural control-flow context.
	
	At the ACFG--FCG level, \toolname{} treats $z_v$ as the function embedding of FCG node $v$. It concatenates $z_v$ with binary-derived structural features describing the function, its neighborhood, local FCG topology, and the whole FCG. The resulting vector is projected into the initial FCG node state $x_v^{(0)}$, after which interprocedural context is propagated over the symmetrized FCG neighborhood:
	\begin{equation}
		x_v^{(k+1)}=\mathrm{ReLU}\left(U_s x_v^{(k)}+
		U_n\cdot \mathrm{Mean}_{u\in \mathcal{N}_{F}(v)} x_u^{(k)}\right),
	\end{equation}
	where $\mathcal{N}_{F}(v)$ denotes the callers and callees adjacent to function $v$ in the symmetrized message-passing graph. The original directed FCG is retained for module traversal.
	
	The resulting representation $x_v^{(L_F)}$ is the context-aware embedding of function node $v$: it combines the intraprocedural function representation learned from $G_v$ with interprocedural calling-context information propagated over the FCG.
	
	\subsection{Anchor Prediction}
	
	Given labeled training binaries, \toolname{} trains the hierarchical encoder as a binary classifier. Let $p(v)$ be the predicted probability that function node $v$ is an anchor, and let $y_v\in\{0,1\}$ be the source-verified label derived in Section~\ref{sec:stable_node_study}. A node is labeled positive when its OSF satisfies the stable boundary condition in Definition~\ref{def:stable_boundary_node}; it is labeled negative when it has verified source semantics but its OSF fails this condition. Nodes whose source semantics cannot be verified are not assigned labels. Let $\Omega$ denote the set of recovered functions with source-verified labels.
	During training, \toolname{} still runs message passing on the full ACFG--FCG, so unlabeled nodes can provide structural context to their labeled neighbors. However, supervision is applied only to nodes in $\Omega$. For each labeled node $v\in\Omega$, the anchor predictor outputs a probability $p_v$, and the model is trained with positive-class weighted binary cross-entropy:
	\begin{equation}
		\mathcal{L}
		=
		-\frac{1}{|\Omega|}
		\sum_{v\in\Omega}
		\left[
		\alpha y_v\log p(v)
		+(1-y_v)\log(1-p(v))
		\right],
	\end{equation}
	where $\alpha=N_{\mathrm{neg}}/N_{\mathrm{pos}}$ compensates for the imbalance between anchor and non-anchor nodes in the training set.
	
	At inference time, \toolname{} predicts $v$ as an anchor when
	\begin{equation}
		p(v) \geq \tau,
	\end{equation}
	where $\tau$ is selected using validation projects and then fixed for held-out test binaries.
	
	The predicted anchors are then used as boundary nodes for the anchor-bounded decomposition described next.
	
	\subsection{Decomposition}
	
	After anchor prediction, \toolname{} decomposes a binary by traversing the FCG from the predicted anchors and the recovered FCG roots, as summarized in Algorithm~\ref{alg:community_partition}. Let $P$ be the predicted anchors and let $R=\{v\in V\mid \deg^{-}(v)=0\}$ be the recovered FCG roots; traversal starts from every node in $P\cup R$. Roots provide entry coverage when a top-level component is not reachable from a predicted anchor. For each node in $P\cup R$, \toolname{} initializes a module at that node and expands along callee edges. Every other predicted anchor serves as a downstream boundary: once such a node is reached, traversal stops before the boundary is included in the current module.
	
	Traversals from nodes in $P\cup R$ are performed independently, so the resulting modules may overlap. This design preserves shared callees and accommodates compiler inlining, through which the same source function may contribute to multiple binary functions and thus to multiple recovered modules. Functions not reached from $P\cup R$ are retained as singleton modules to ensure full FCG coverage. Thus, \toolname{} produces an overlapping, coverage-preserving module decomposition rather than a strict partition.
	
	\newcommand{\INPUT}{\item[\textbf{Input:}]}
	\newcommand{\OUTPUT}{\item[\textbf{Output:}]}
	\renewcommand{\algorithmiccomment}[1]{\hfill{\footnotesize\textit{// #1}}}
	
	\begin{algorithm}[t]
		\caption{Anchor-bounded FCG decomposition.}
		\label{alg:community_partition}
		\begin{algorithmic}[1]
			\INPUT function call graph $G=(V,E)$, predicted anchor set $P \subseteq V$
			\OUTPUT anchor-bounded module set $M$
			
			\STATE $M \gets \emptyset$
			\STATE $R \gets \{v\in V\mid \deg^{-}(v)=0\}$
			\STATE $C \gets P \cup R$
			\STATE $Q \gets C$
			\WHILE{$Q$ is not empty}
			\STATE $c \gets \text{dequeue}(Q)$
			\STATE $m \gets \{c\}$
			\STATE $S \gets \text{successors}(c,G)$
			\WHILE{$S$ is not empty}
			\STATE $v \gets \text{dequeue}(S)$
			\IF{$v \notin C$ \AND $v \notin m$}
			\STATE $m \gets m \cup \{v\}$
			\STATE $S \gets S \cup \text{successors}(v,G)$
			\ENDIF
			\ENDWHILE
			\STATE $M \gets M \cup \{m\}$
			\ENDWHILE
			\STATE $V_{\mathrm{covered}} \gets \bigcup_{m\in M}m$
			\FOR{each $v\in V\setminus V_{\mathrm{covered}}$}
			\STATE $M \gets M \cup \{\{v\}\}$
			\ENDFOR
			\RETURN $M$
		\end{algorithmic}
	\end{algorithm}
	\vspace{-5pt}
	
	\subsection{Module-Level Matching}
	
	Given a query binary and a candidate binary, \toolname{} first applies the above decomposition procedure to obtain two module sets, denoted as $\mathcal{M}_q$ and $\mathcal{M}_c$. The task of module-level matching is to rank candidate modules in $\mathcal{M}_c$ for each query module $m_q\in\mathcal{M}_q$. 
	
	\toolname{} follows the module matching framework of ModX, which compares two modules using the three components shown in Figure~\ref{fig:overview}: \textit{Syntactic Similarity}, \textit{Semantic Similarity}, and \textit{Graph Similarity}. Syntactic Similarity is derived from string literals and constant values extracted from the modules. Graph Similarity captures module call-graph topology using graph-kernel and edge-topology information.
	
	The main difference between \toolname{} and ModX lies in how Semantic Similarity is computed. ModX compares function-level representations and aggregates selected function-pair similarities into a module-level score. This function-level aggregation becomes fragile under function inlining, where the granularity of recovered binary functions may differ across compilation variants. For example, a function in one binary may contain the inlined code of several source functions, while the corresponding code remains distributed across multiple functions in another binary. In such cases, one-to-one function comparisons can underestimate the similarity between two modules that still cover related source-level behavior.
	
	Let $e_v$ denote the semantic embedding of a recovered binary function $v$, obtained from a pretrained binary function encoder. For each module $m$, \toolname{} constructs the directed FCG subgraph induced by its member functions and uses their function embeddings as node features. Each graph-encoding layer aggregates caller and callee representations separately, combines them with the current node state, and applies a residual connection followed by normalization. Gated attention pooling, together with a mean-pooled residual, then produces a normalized module graph embedding $g(m)$. Given two modules $m_q$ and $m_c$, their Semantic Similarity is the cosine similarity between $g(m_q)$ and $g(m_c)$.

	Training uses source mappings only to construct contrastive supervision. Let $\mathcal{S}(m)$ denote the set of source functions mapped to module $m$. We define their source-function Dice as
	\begin{equation}
		D(m_q,m_c)=
		\frac{2|\mathcal{S}(m_q)\cap \mathcal{S}(m_c)|}
		{|\mathcal{S}(m_q)|+|\mathcal{S}(m_c)|}.
	\end{equation}
	Candidates with $D(m_q,m_c)>0$ are treated as positives and weighted by their Dice scores, whereas candidates with zero source overlap are treated as negatives. 
	
	Among candidates with zero source overlap, we prioritize those receiving high scores from the fixed combination of Syntactic Similarity, Graph Similarity, and the function-level semantic similarity used by ModX. These hard negatives resemble the query according to binary-only evidence and therefore provide more informative contrastive supervision. The resulting multi-positive contrastive objective brings source-overlapping modules closer while separating non-overlapping modules. Source mappings are not used to construct embeddings or rank candidates at inference time.
	
	Given a query module $m_q$ and a candidate module $m_c$, \toolname{} computes their matching score by combining Syntactic Similarity, Semantic Similarity, and Graph Similarity:
	\begin{equation}
		\begin{aligned}
			S(m_q,m_c) &=
			\alpha \mathrm{Sim}_{\mathrm{syn}}(m_q,m_c)
			+ \beta \mathrm{Sim}_{\mathrm{sem}}(m_q,m_c) \\
			&\quad
			+ \gamma \mathrm{Sim}_{\mathrm{graph}}(m_q,m_c).
		\end{aligned}
	\end{equation}
	Here, $\mathrm{Sim}_{\mathrm{syn}}$, $\mathrm{Sim}_{\mathrm{sem}}$, and $\mathrm{Sim}_{\mathrm{graph}}$ denote Syntactic Similarity, Semantic Similarity, and Graph Similarity, respectively. Syntactic Similarity and Graph Similarity follow the ModX implementations, whereas Semantic Similarity is computed from the contrastively learned module graph embeddings. The three components are combined using a configuration selected on a disjoint validation set and fixed before testing.
	
	For each query module, \toolname{} ranks candidate modules according to their matching scores. Because the semantic representation jointly models function embeddings and module-internal call structure, the comparison does not require one-to-one correspondence between recovered functions and is therefore less sensitive to function-granularity changes introduced by inlining.
	
	\section{Evaluation}
	\label{sec:evaluation}
	
	This section evaluates \toolname{} on binary decomposition and module-level matching under cross-compilation settings. We first compare \toolname{} with existing methods on BinKit and then assess its effectiveness in third-party component reuse scenarios derived from ISRD~\cite{ISRD}. We further study how the main design choices affect the results and measure the runtime overhead of the framework. The evaluation answers the following questions:
	
	\begin{itemize}
		\item \textbf{RQ1.} How effective is \toolname{} in binary decomposition and module matching compared with existing methods?
		\item \textbf{RQ2.} How effective is \toolname{} in identifying reused modules in third-party component reuse scenarios?
		\item \textbf{RQ3.} How do the anchor prediction model and the implementation of Semantic Similarity affect \toolname{}?
		\item \textbf{RQ4.} What is the runtime overhead of \toolname{} compared with existing decomposition and matching baselines?
	\end{itemize}
	
	\subsection{Experiment Setup}
	\label{sec:setup}
	
	\subsubsection{Datasets and Splitting}
	
	\textbf{BinKit.} We conduct the main decomposition and matching experiments on the 3,024 x86-64 binaries from 84 executable targets and 28 open-source projects described in Section~\ref{sec:stable_node_study}. Source-grounded labels and evaluation oracles are derived from the corresponding binaries with debug information and transferred to their stripped counterparts.
	
	To avoid project leakage, we use project-level ten-fold cross-validation. For each held-out test fold, eight folds are used for training and one for validation. Anchor-prediction and module-matching configurations are selected only on the validation fold and fixed before testing. For fair comparison, \toolname{} and all baselines are evaluated on the same stripped binaries.

	\textbf{ISRD.} To assess generalization beyond BinKit in RQ2, we use a regenerated LTO dataset aligned with the binaries retained by ISRD~\cite{ISRD}. The corpus contains 608 x86-64 stripped binaries from 26 source projects, compiled with GCC~8.2.0 and Clang~7.0 at optimization levels O0--O3. Ground truth is constructed from source-function correspondences within the original ISRD reuse relations, as detailed in RQ2. We retain 864 debug-line-verified positive binary pairs. \toolname{} is trained on BinKit and transferred to ISRD without tuning.
	
	\subsubsection{Baselines}
	We compare \toolname{} with two binary decomposition baselines. ModX~\cite{yang2022modx} is a clustering-based decomposition method that groups functions according to FCG structure and supports module matching with Syntactic Similarity, Semantic Similarity, and Graph Similarity. BMVul~\cite{guo2023searching} decomposes binaries with graph community detection and is used as a decomposition baseline. Since BMVul does not define a downstream module matching stage comparable to ModX, it is included only in the binary decomposition comparison in RQ1. ModX is used in both the BinKit module matching comparison and the ISRD external validation.
	
	\subsubsection{Evaluation Metrics}
	
	We use separate metrics for anchor prediction, decomposition quality, and module matching. Anchor prediction metrics evaluate the identification of binary functions that serve as stable boundaries, decomposition metrics measure whether generated modules preserve source-level semantics across compilation variants, and module matching metrics evaluate whether these modules can retrieve semantically corresponding modules from another stripped binary.

	\textbf{Anchor prediction metrics.} Anchor prediction is evaluated over source-verified binary functions represented as FCG nodes. Binary functions labeled as source-verified stable boundaries are positives, other source-verified binary functions are negatives, and unlabeled FCG nodes are excluded. Let $TP$, $FP$, and $FN$ denote the numbers of true positives, false positives, and false negatives, respectively. We report precision, recall, and F1:
	\begin{equation}
		\mathrm{Precision}=\frac{TP}{TP+FP}.
	\end{equation}
	\begin{equation}
		\mathrm{Recall}=\frac{TP}{TP+FN}.
	\end{equation}
	\begin{equation}
		\mathrm{F1}=\frac{2\,\mathrm{Precision}\,\mathrm{Recall}}
		{\mathrm{Precision}+\mathrm{Recall}}.
	\end{equation}
	
	\textbf{Decomposition metrics.} We use \textit{module decomposition quality} (MDQ) to measure how well a generated module aligns with modules in another binary. For a module $m$, let $V(m)$ be its binary-function set and let $\mu(f)$ be the source functions mapped from binary function $f$. For a query module $m_q$ and a reference module $m_r$, we define $S_q=\bigcup_{f\in V(m_q)}\mu(f)$, with $S_r$ defined analogously. Thus, $S_q$ and $S_r$ contain the distinct mapped source functions of the two modules. At the source-function level, module overlap is measured using the S{\o}rensen--Dice similarity:
	\begin{equation}
		Sim_{\mathrm{func}}(m_q,m_r)=
		\frac{2|S_q \cap S_r|}{|S_q|+|S_r|}.
	\end{equation}
	
	To account for function size, we also define a function-length-weighted similarity, where each source function is weighted by its source-line length $\ell(s)$:
	\begin{equation}
		Sim_{\mathrm{len}}(m_q,m_r)=
		\frac{2\sum_{s\in S_q\cap S_r}\ell(s)}
		{\sum_{s\in S_q}\ell(s)+\sum_{s\in S_r}\ell(s)}.
	\end{equation}
	Here, $|S_q\cap S_r|$ is the number of mapped source functions shared by the two modules, while $\sum_{s\in S_q\cap S_r}\ell(s)$ is their total source-line length. The denominators normalize these shared quantities by the combined function count or source-line length of the two modules. Both similarities range from 0 to 1: 0 indicates that the modules share no mapped source function, whereas 1 indicates identical mapped source-function content under the corresponding weighting.
	
	Let $\mathcal{M}_q$ and $\mathcal{M}_r$ denote the sets of modules recovered from the query and reference binaries, respectively. For either similarity measure, MDQ is defined as
	\begin{equation}
		\mathrm{MDQ}_{\tau}(m_q)
		=\max_{m_r\in\mathcal{M}_r} Sim_{\tau}(m_q,m_r),
		\quad \tau\in\{\mathrm{func},\mathrm{len}\}.
	\end{equation}
	Thus, $\mathrm{MDQ}_{\tau}(m_q)$ is the highest source-level overlap between $m_q$ and any module in the reference binary. It measures the best achievable source-level alignment under a given decomposition, independent of any learned module matcher. Each binary pair is evaluated directionally, so both $A\rightarrow B$ and $B\rightarrow A$ are counted.

	Query modules with $S_q=\emptyset$ are excluded because their source-level overlap cannot be evaluated. We define that a query module is \textit{overlap-eligible} if $\mathrm{MDQ}_{\mathrm{func}}(q)>0$. The set of all overlap-eligible queries is therefore
	\begin{equation}
		\mathcal{Q}^{+}
		=\{q\in\mathcal{M}_q\mid \mathrm{MDQ}_{\mathrm{func}}(q)>0\}.
	\end{equation}
	All decomposition statistics reported below (\textbf{Avg MDQ}, \textbf{Perfect}, and \textbf{Avg Size}) are computed over $\mathcal{Q}^{+}$. Avg MDQ is the average MDQ, Perfect is the proportion of queries with MDQ equal to 1.0, and Avg Size is the average number of recovered binary functions per query module.
	
	\textbf{Module matching metrics.} For downstream module matching, candidate modules are ranked using binary-derived module similarities. Because $\mathcal{Q}^{+}$ is defined using MDQ over the complete reference set $\mathcal{M}_r$, query eligibility is independent of the ranked candidate list produced by the matcher. For each $q\in\mathcal{Q}^{+}$, let $(c_{q,1},c_{q,2},\ldots)$ be its ranked candidate list. \textit{Top-1 Similarity} measures the source-function Dice between $q$ and the top-ranked candidate:
	\begin{equation}
		\mathrm{Top1Sim}
		=
		\frac{1}{|\mathcal{Q}^{+}|}
		\sum_{q\in\mathcal{Q}^{+}}
		Sim_{\mathrm{func}}(q,c_{q,1}).
	\end{equation}
	Recall@$k$ measures whether the top-$k$ list contains a source-overlapping candidate:
	\begin{equation}
		\mathrm{Recall@}k=\frac{1}{|\mathcal{Q}^{+}|}
		\sum_{q\in\mathcal{Q}^{+}}
		\mathbb{I}\left[
		\exists j\le k,\ Sim_{\mathrm{func}}(q,c_{q,j})>0
		\right].
	\end{equation}
	We also report mean reciprocal rank (MRR), computed from the rank of the first candidate $c_{q,j}$ satisfying $Sim_{\mathrm{func}}(q,c_{q,j})>0$ and averaged over the same query set. Thus, Top-1 Similarity, Recall@$k$, and MRR are all computed over overlap-eligible queries. All reported test point estimates are pooled at the query level. Avg MDQ, defined above, characterizes the decomposition-imposed upper bound independently of the learned ranking.
	
\subsubsection{Implementation Details}

During graph construction, we use IDA Pro~\cite{IDAPro} to recover FCGs and ACFGs, Capstone~\cite{capstone} to identify direct cross-function jumps, and Readelf~\cite{readelf} to obtain supervision and evaluation mappings. Compiler, optimization, and symbol-name metadata are not used as model inputs.

For anchor prediction, the hierarchical ACFG--FCG encoder is trained with AdamW. Predicted anchors and FCG roots are then processed by the decomposition procedure described above.

For module matching, we use the released Gemini checkpoint~\cite{Gemini} in inference-only mode to obtain function embeddings from the IDA-derived ACFGs. The module graph encoder contains two directed message-passing layers followed by gated attention pooling with a mean residual. The encoder and matching weights $(\alpha,\beta,\gamma)$ are selected exclusively on the validation folds according to mean MRR. This yields $\alpha=0.475$, $\beta=0.350$, and $\gamma=0.175$ for Syntactic Similarity, Semantic Similarity, and Graph Similarity, respectively. All selected configurations are frozen before testing.

All experiments are conducted on a Windows 11 workstation with an Intel Core Ultra 9 185H processor, 32~GB of memory, and an NVIDIA GeForce RTX 4070 GPU with 8~GB of memory.
	
	\subsection{RQ1: Comparison with Existing Methods on BinKit}
	\label{sec:binkit_effectiveness}
	
	RQ1 evaluates \toolname{} on BinKit at both stages of the pipeline: binary decomposition and module matching.

	\subsubsection{Binary Decomposition}
	\label{sec:tool_binary_decomposition}
	
Before evaluating decomposition quality, we first examine how the validation-selected anchor threshold affects anchor prediction against source-verified labels.

Table~\ref{tab:anchor_threshold_ablation} reports anchor prediction precision, recall, and F1 under different thresholds. Lower thresholds preserve higher recall, whereas higher thresholds improve precision. We select 0.4 because it achieves the highest validation F1.

	\begin{table}[t]
		\centering
		\caption{Anchor prediction under different thresholds.}
		\label{tab:anchor_threshold_ablation}
		\footnotesize
		\begin{tabular}{c|ccc}
			\hline
			Threshold & Precision & Recall & F1 \\ \hline
			0.10 & 0.486 & 0.939 & 0.640 \\
			0.20 & 0.536 & 0.900 & 0.672 \\
			0.30 & 0.575 & 0.851 & 0.687 \\
			\textbf{0.40} & \textbf{0.611} & \textbf{0.797} & \textbf{0.692} \\
			0.50 & 0.640 & 0.724 & 0.679 \\
			0.60 & 0.671 & 0.605 & 0.636 \\
			0.70 & 0.710 & 0.475 & 0.570 \\
			0.80 & 0.765 & 0.327 & 0.458 \\
			0.90 & 0.828 & 0.213 & 0.339 \\
			1.00 & 1.000 & 0.001 & 0.002 \\ \hline
		\end{tabular}
	\end{table}

Table~\ref{tab:main_decomposition_results} compares \toolname{}, ModX, and BMVul on binary decomposition across the three cross-compilation settings. \toolname{} and BMVul produce overlap-allowed regions, whereas ModX produces an exclusive partition.
	
	\begin{table*}[t]
		\centering
		\caption{Binary decomposition evaluation on BinKit.}
		\label{tab:main_decomposition_results}
		\footnotesize
		\begin{tabular}{c|c|c|c|c|c}
			\hline
			Setting & Method & Avg Size & Avg MDQ (Func.) & Avg MDQ (Len.-Wtd.) & Perfect \\ \hline
			\multirow{3}{*}{Cross-Compiler, Same-Optimization}
			& \toolname{} & 6.34 & \textbf{0.807} & \textbf{0.800} & 33.85\% \\
			& ModX & 1.50 & 0.725 & 0.733 & 43.82\% \\
			& BMVul & 8.30 & 0.780 & 0.783 & \textbf{64.44\%} \\ \hline
			\multirow{3}{*}{Same-Compiler, Cross-Optimization}
			& \toolname{} & 6.42 & \textbf{0.809} & \textbf{0.801} & 40.80\% \\
			& ModX & 1.50 & 0.676 & 0.706 & 41.20\% \\
			& BMVul & 8.89 & 0.673 & 0.686 & \textbf{52.11\%} \\ \hline
			\multirow{3}{*}{Cross-Compiler, Cross-Optimization}
			& \toolname{} & 6.72 & \textbf{0.759} & \textbf{0.748} & 28.05\% \\
			& ModX & 1.52 & 0.592 & 0.620 & 27.83\% \\
			& BMVul & 9.60 & 0.589 & 0.602 & \textbf{41.31\%} \\ \hline
			\multirow{3}{*}{Overall}
			& \toolname{} & 6.49 & \textbf{0.793} & \textbf{0.784} & 34.34\% \\
			& ModX & 1.50 & 0.667 & 0.689 & 37.93\% \\
			& BMVul & 8.88 & 0.687 & 0.697 & \textbf{53.43\%} \\ \hline
		\end{tabular}
	\end{table*}
	
	\toolname{} achieves the highest Avg MDQ in all three settings under both source-function and function-length-weighted evaluation. Its overall source-function Avg MDQ is 0.793, compared with 0.667 for ModX and 0.687 for BMVul. BMVul has the highest Perfect ratio, 53.43\%, but its overall Avg MDQ remains below \toolname{}, showing that more exact matches do not yield stronger average alignment across all evaluated queries. This consistent advantage indicates that anchor-bounded decomposition preserves source-level alignment across compilation variants more effectively than the baseline decompositions.
	
	\subsubsection{Module Matching}
	\label{sec:module_matching_eval}
	
Table~\ref{tab:module_matching_results} compares \toolname{} and ModX in an end-to-end module matching experiment over 3,000 sampled stripped-binary pairs, with each method using its own decomposition output.
	
	\begin{table*}[t]
		\centering
		\footnotesize
		\caption{Module matching evaluation on BinKit.}
		\label{tab:module_matching_results}
		\begin{tabular}{c|c|c|c|c|c}
			\hline
			Setting & Method & Top-1 Sim. & Recall@1 & Recall@5 & MRR \\ \hline
			\multirow{2}{*}{Cross-Compiler, Same-Optimization}
			& \toolname{} & \textbf{0.587} & \textbf{0.798} & \textbf{0.888} & \textbf{0.839} \\
			& ModX & 0.401 & 0.523 & 0.673 & 0.594 \\ \hline
			\multirow{2}{*}{Same-Compiler, Cross-Optimization}
			& \toolname{} & \textbf{0.601} & \textbf{0.798} & \textbf{0.879} & \textbf{0.835} \\
			& ModX & 0.412 & 0.522 & 0.631 & 0.577 \\ \hline
			\multirow{2}{*}{Cross-Compiler, Cross-Optimization}
			& \toolname{} & \textbf{0.493} & \textbf{0.748} & \textbf{0.843} & \textbf{0.792} \\
			& ModX & 0.272 & 0.407 & 0.535 & 0.472 \\ \hline
			\multirow{2}{*}{Overall}
			& \toolname{} & \textbf{0.562} & \textbf{0.782} & \textbf{0.871} & \textbf{0.823} \\
			& ModX & 0.364 & 0.487 & 0.616 & 0.550 \\ \hline
		\end{tabular}
	\end{table*}
	
	Overall, \toolname{} outperforms ModX on all four metrics: Top-1 Similarity is 0.562 versus 0.364, Recall@1 is 0.782 versus 0.487, Recall@5 is 0.871 versus 0.616, and MRR is 0.823 versus 0.550. This consistent end-to-end advantage indicates that combining anchor-bounded decomposition with module graph embeddings improves module retrieval over the ModX pipeline.
	
	For \toolname{}, the cross-compiler and cross-optimization settings yield similar results, with Top-1 Similarity values of 0.587 and 0.601 and MRR values of 0.839 and 0.835. When both compiler and optimization differ, these metrics decrease to 0.493 and 0.792, respectively. This indicates that \toolname{} is robust when either the compiler or
	the optimization level changes, although simultaneous changes in both
	dimensions make module matching more challenging. 

	\subsection{RQ2: Third-Party Component Reuse}
	\label{sec:isrd_external_validation}
	
	RQ2 examines whether \toolname{} generalizes from BinKit to cross-project component reuse on ISRD.

	Table~\ref{tab:isrd_module_matching_results} compares \toolname{} and ModX using 16,166 cross-project source-function correspondences as ground truth. Of the 864 positive binary pairs, 562 yield at least one query module with a source-corresponding candidate in each direction and therefore enter the ranking evaluation.

	\begin{table*}[t]
		\centering
		\footnotesize
		\caption{Module retrieval evaluation on ISRD.}
		\label{tab:isrd_module_matching_results}
		\begin{tabular}{c|c|c|c|c|c}
			\hline
			Setting & Method & Top-1 Sim. & Recall@1 & Recall@5 & MRR \\ \hline
			\multirow{2}{*}{Cross-Compiler, Same-Optimization}
			& \toolname{} & \textbf{0.280} & \textbf{0.482} & \textbf{0.679} & \textbf{0.577} \\
			& ModX & 0.249 & 0.368 & 0.615 & 0.486 \\ \hline
			\multirow{2}{*}{Same-Compiler, Cross-Optimization}
			& \toolname{} & \textbf{0.225} & \textbf{0.442} & \textbf{0.631} & \textbf{0.532} \\
			& ModX & 0.174 & 0.272 & 0.502 & 0.385 \\ \hline
			\multirow{2}{*}{Cross-Compiler, Cross-Optimization}
			& \toolname{} & \textbf{0.190} & \textbf{0.409} & \textbf{0.641} & \textbf{0.514} \\
			& ModX & 0.120 & 0.223 & 0.465 & 0.342 \\ \hline
			\multirow{2}{*}{Overall}
			& \toolname{} & \textbf{0.231} & \textbf{0.444} & \textbf{0.650} & \textbf{0.541} \\
			& ModX & 0.182 & 0.288 & 0.528 & 0.405 \\ \hline
		\end{tabular}
	\end{table*}

	Overall, \toolname{} outperforms ModX on all four metrics: Top-1 Similarity is 0.231 versus 0.182, Recall@1 is 0.444 versus 0.288, Recall@5 is 0.650 versus 0.528, and MRR is 0.541 versus 0.405. Across the 562 evaluated pairs, all four paired improvements are statistically significant, indicating that \toolname{} more accurately localizes reused modules than ModX in the cross-project setting.

	Nevertheless, the absolute Top-1 Similarity and MRR of \toolname{} are lower than on BinKit, where they reach 0.562 and 0.823. BinKit compares compilation variants of the same target, whereas ISRD matches reused source regions embedded in different projects, whose surrounding functions and call structures may differ substantially. Moreover, the model and matching weights are selected on BinKit and transferred to ISRD without dataset-specific tuning. The lower absolute results therefore reflect both the greater difficulty of cross-project partial reuse and the resulting domain shift.
	
	\subsection{RQ3: Ablation Study}
	\label{sec:ablation_study}
	
	RQ3 uses controlled ablations to isolate the contributions of hierarchical anchor prediction and module graph embeddings.

	Table~\ref{tab:anchor_predictor_ablation} compares the hierarchical anchor predictor with ACFG-only and FCG-only variants. Table~\ref{tab:module_matching_ablation} compares module graph and function-level embeddings with the \toolname{} decomposition fixed.

	\begin{table}[t]
		\centering
		\footnotesize
		\setlength{\tabcolsep}{3pt}
		\caption{Anchor-predictor ablation.}
		\label{tab:anchor_predictor_ablation}
		\begin{tabular}{l|ccc|c}
			\hline
			\multirow{2}{*}{Configuration} &
			\multicolumn{3}{c|}{Anchor Prediction} &
			\multicolumn{1}{c}{Decomposition} \\ \cline{2-5}
			& Precision & Recall & F1 & Avg MDQ \\ \hline
			Hierarchical ACFG--FCG & \textbf{0.611} & 0.797 & \textbf{0.692} & \textbf{0.793} \\
			ACFG only & 0.460 & 0.680 & 0.549 & 0.665 \\
			FCG only & 0.550 & \textbf{0.858} & 0.670 & 0.786 \\ \hline
		\end{tabular}
	\end{table}

	\begin{table}[t]
		\centering
		\footnotesize
		\setlength{\tabcolsep}{3pt}
		\caption{Semantic representation ablation.}
		\label{tab:module_matching_ablation}
		\begin{tabular}{l|cccc}
			\hline
			Configuration & Top-1 Sim. & Recall@1 & Recall@5 & MRR \\ \hline
			Module graph embedding & \textbf{0.5621} & \textbf{0.7821} & \textbf{0.8708} & \textbf{0.8228} \\
			Function-level embedding & 0.5515 & 0.7715 & 0.8634 & 0.8137 \\ \hline
		\end{tabular}
	\end{table}

	The hierarchical predictor achieves the highest anchor F1 (0.692) and Avg MDQ (0.793). Removing FCG propagation reduces these values to 0.549 and 0.665. Relative to FCG-only, adding ACFG evidence raises precision from 0.550 to 0.611 and F1 from 0.670 to 0.692, despite reducing recall from 0.858 to 0.797. Together, these results support combining intraprocedural ACFG evidence with interprocedural FCG context for anchor prediction.

	With decomposition fixed, the module graph embedding outperforms the function-level embedding on all four metrics, increasing Top-1 Similarity from 0.5515 to 0.5621 and MRR from 0.8137 to 0.8228. The module graph embedding improves MRR in all ten held-out test folds, and the improvement remains statistically significant after Holm correction (adjusted $p=0.0078$). This controlled comparison supports using module-internal FCG structure rather than function embeddings alone.

	\subsection{RQ4: Runtime Overhead}
	\label{sec:runtime_overhead}
	
	RQ4 measures how the decomposition and module-matching runtimes scale with FCG size.
	
\begin{figure}[t]
		\centering
		\includegraphics[width=0.98\columnwidth]{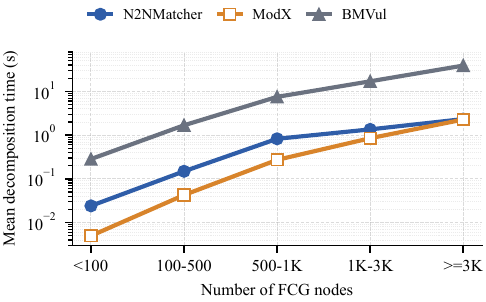}
		\caption{Decomposition runtime by FCG size.}
		\label{fig:rq5_decomposition_runtime}
	\end{figure}

	Figure~\ref{fig:rq5_decomposition_runtime} compares decomposition runtime from prebuilt FCGs and ACFGs to generated modules across five FCG-size ranges. The measurement includes all processing within this boundary: anchor prediction and anchor-bounded decomposition for \toolname{}, and the corresponding decomposition procedures for ModX and BMVul. Disassembly, FCG and ACFG construction, and training are excluded. All measurements are executed serially, and each binary is measured once. 

\toolname{} is slower than ModX on small binaries, but the gap narrows with FCG size and reaches 1.02$\times$ at 3,000 nodes or more. It remains 9.1--16.7$\times$ faster than BMVul, indicating that anchor prediction adds limited decomposition overhead at larger scales.

\begin{figure}[t]
		\centering
		\includegraphics[width=0.98\columnwidth]{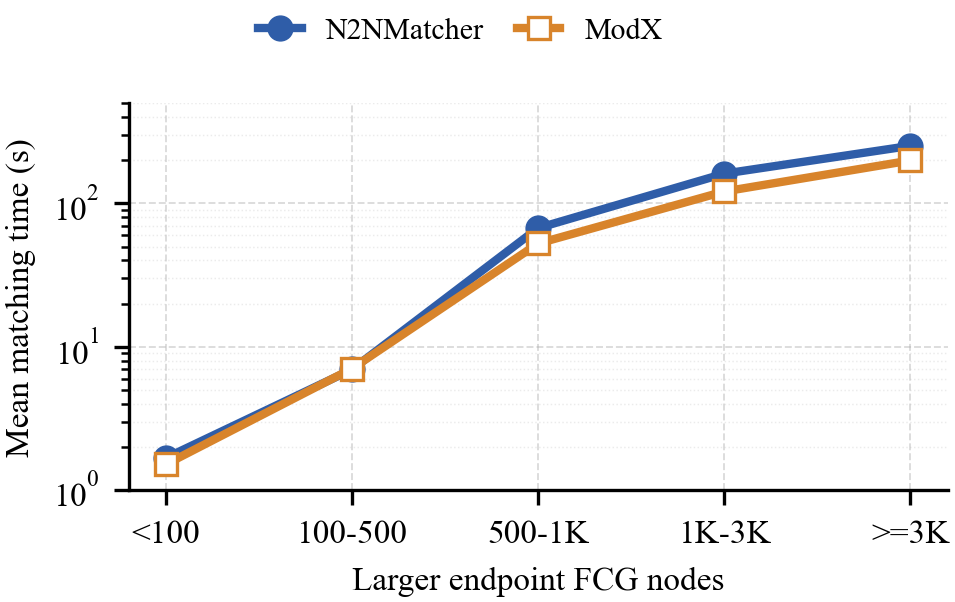}
		\caption{Module matching runtime by FCG size.}
		\label{fig:rq5_matching_runtime}
	\end{figure}

Figure~\ref{fig:rq5_matching_runtime} compares end-to-end matching runtime across five ranges defined by the larger FCG in each binary pair. Timing starts from each method's generated modules and the prebuilt FCGs and ACFGs, and ends with ranked candidate modules. For both methods, it includes Gemini inference for function embeddings, the computation of Syntactic, Semantic, and Graph Similarity, and module ranking; \toolname{} additionally includes loading and applying the module graph encoder. 

Matching time is comparable below 500 nodes. Across the complete batch, \toolname{} requires 4.01 hours versus 3.12 hours for ModX, an overhead of 28.3\%. Applying the module graph encoder adds only 13.0 seconds; most of the additional cost occurs during module ranking.
	
	\section{Discussion}
	\label{sec:discussion}
	
	\textbf{FCG changes caused by optimizations other than inlining.} Several compiler transformations besides inlining can alter recovered FCGs. Dead-code and unreachable-function elimination remove nodes and incident edges; function cloning, specialization, and interprocedural constant propagation may duplicate functions or eliminate calls; function merging and code folding may collapse distinct functions; outlining and function splitting may introduce additional nodes; and tail-call conversion or indirect-call resolution may redirect edges. Inlining nevertheless remains a major measured contributor: 217,649 of 855,226 source-mapped binary functions (25.45\%) contain mappings to multiple source functions, and this proportion rises from 0.25\% at O0 to 23.87\%, 46.46\%, and 63.97\% at O1, O2, and O3, respectively. Our source-grounded anchor labeling remains applicable under these transformations because it relies on binary-to-source mappings across configurations rather than unchanged FCG topology. A source function need not be recovered in every configuration; where it is recovered as an OSF and it is never observed as inlined under another OSF, it is labeled as an anchor. Thus, node or edge removal, duplication, merging, splitting, and redirection do not by themselves invalidate the labeling procedure. 

	\textbf{Performance degradation on ISRD and potential improvements.} \toolname{} achieves lower absolute retrieval scores on ISRD than on BinKit because the two datasets represent different matching conditions. BinKit compares compilation variants of the same target, whereas ISRD contains partial reuse across different projects, so reused functions may be surrounded by unrelated code and participate in different call structures. Moreover, the trained module graph encoder and matching weights are selected on BinKit and transferred to ISRD without cross-project adaptation. Future work could train the module encoder with project-disjoint cross-project reuse pairs, use hard negatives from unrelated projects, and select matching weights on a separate cross-project validation set. Partial-graph alignment or attention that suppresses project-specific context may further improve the localization of modified reused regions.

	\section{Threats to Validity}
	
	\textbf{Internal validity.} Errors in recovered function boundaries, call edges, basic blocks, or instructions may affect the results~\cite{pang2021sok}. We use the same IDA Pro~\cite{IDAPro} pipeline for all methods, but different methods may respond differently to these errors. 
	
	\textbf{External validity.} The evaluation covers x86-64 LTO binaries from BinKit and ISRD compiled with GCC or Clang. Results may differ for other architectures, toolchains, non-LTO or obfuscated binaries, and open-world component detection.
	
	\textbf{Construct validity.} Training labels and relevance judgments rely on debug-line mappings, which may be incomplete under optimization and may not capture all forms of semantic reuse. The conservative ISRD oracle may miss modified functions, and Avg MDQ may favor coarse modules. We therefore report Perfect and Avg Size as complementary decomposition metrics.
	
	\section{Conclusion}
	
	We presented \toolname{}, an inlining-resilient framework for binary decomposition and module matching. Our source-grounded study shows that compilation variance disrupts FCG structure and function-level alignment while leaving stable boundary nodes. Based on this finding, we designed \toolname{} to predict these boundaries, construct anchor-bounded modules, and match them with module graph embeddings. Our experiments show that it outperforms ModX and BMVul in decomposition and ModX in module matching on BinKit, and raises Recall@1 over ModX from 0.288 to 0.444 on ISRD.
	
	In future work, we will improve \toolname{}'s ability to identify reused modules under more challenging conditions, including unseen projects, diverse architectures and toolchains, and open-world reuse scenarios. We hope that our study could provide useful insights for future research on program-level binary analysis.

	\bibliographystyle{IEEEtran}
	\bibliography{reference}

@misc{CodeReuse,
	title = {{2025 Open Source Security and Risk Analysis Report}},
	howpublished = "\url{https://www.blackduck.com/content/dam/black-duck/en-us/reports/rep-ossra.pdf}",
	year = {2025}, 
	note = "[Online; accessed 9-August-2026]"
}

@misc{zlib,
	title = {{madler/zlib GitHub Repository}},
	howpublished = "\url{https://github.com/madler/zlib}",
	year = {2026}, 
	note = "[Online; accessed 9-August-2026]"
}

@misc{curl,
	title = {{curl/curl GitHub Repository}},
	howpublished = "\url{https://github.com/curl/curl}",
	year = {2026}, 
	note = "[Online; accessed 9-August-2026]"
}

@article{gkortzis2021software,
	title={Software reuse cuts both ways: An empirical analysis of its relationship with security vulnerabilities},
	author={Gkortzis, Antonios and Feitosa, Daniel and Spinellis, Diomidis},
	journal={Journal of Systems and Software},
	volume={172},
	pages={110653},
	year={2021},
	publisher={Elsevier},
	doi={10.1016/j.jss.2020.110653}
}

@article{kula2018developers,
	title={Do developers update their library dependencies? An empirical study on the impact of security advisories on library migration},
	author={Kula, Raula Gaikovina and German, Daniel M and Ouni, Ali and Ishio, Takashi and Inoue, Katsuro},
	journal={Empirical Software Engineering},
	volume={23},
	number={1},
	pages={384--417},
	year={2018},
	publisher={Springer},
	doi={10.1007/s10664-017-9521-5}
}

@article{li2023libam,
	title={Libam: An area matching framework for detecting third-party libraries in binaries},
	author={Li, Siyuan and Wang, Yongpan and Dong, Chaopeng and Yang, Shouguo and Li, Hong and Sun, Hao and Lang, Zhe and Chen, Zuxin and Wang, Weijie and Zhu, Hongsong and others},
	journal={ACM Transactions on Software Engineering and Methodology},
	volume={33},
	number={2},
	pages={1--35},
	year={2023},
	publisher={ACM New York, NY},
	doi={10.1145/3625294}
}

@inproceedings{dong2024libvdiff,
	title={LibvDiff: Library Version Difference Guided OSS Version Identification in Binaries},
	author={Dong, Chaopeng and Li, Siyuan and Yang, Shouguo and Xiao, Yang and Wang, Yongpan and Li, Hong and Li, Zhi and Sun, Limin},
	booktitle={Proceedings of the 46th IEEE/ACM International Conference on Software Engineering},
	pages={1--12},
	year={2024},
	doi={10.1145/3597503.3623336}
}

@inproceedings{tang2022libdb,
	title={Libdb: An effective and efficient framework for detecting third-party libraries in binaries},
	author={Tang, Wei and Wang, Yanlin and Zhang, Hongyu and Han, Shi and Luo, Ping and Zhang, Dongmei},
	booktitle={Proceedings of the 19th International Conference on Mining Software Repositories},
	pages={423--434},
	year={2022},
	doi={10.1145/3524842.3528442}
}

@article{zhu2022bbdetector,
	title={BBDetector: A precise and scalable third-party library detection in binary executables with fine-grained function-level features},
	author={Zhu, Xiaoya and Wang, Junfeng and Fang, Zhiyang and Yin, Xiaokang and Liu, Shengli},
	journal={Applied Sciences},
	volume={13},
	number={1},
	pages={413},
	year={2022},
	publisher={MDPI},
	doi={10.3390/app13010413}
}

@inproceedings{yang2022modx,
	title={ModX: binary level partially imported third-party library detection via program modularization and semantic matching},
	author={Yang, Can and Xu, Zhengzi and Chen, Hongxu and Liu, Yang and Gong, Xiaorui and Liu, Baoxu},
	booktitle={Proceedings of the 44th International Conference on Software Engineering},
	pages={1393--1405},
	year={2022},
	doi={10.1145/3510003.3510627}
}

@article{sun2023moddiff,
	title={ModDiff: modularity similarity-based malware homologation detection},
	author={Sun, Huaqi and Shu, Hui and Kang, Fei and Guang, Yan},
	journal={Electronics},
	volume={12},
	number={10},
	pages={2258},
	year={2023},
	publisher={MDPI},
	doi={10.3390/electronics12102258}
}

@inproceedings{karande2018bcd,
	title={BCD: Decomposing binary code into components using graph-based clustering},
	author={Karande, Vishal and Chandra, Swarup and Lin, Zhiqiang and Caballero, Juan and Khan, Latifur and Hamlen, Kevin},
	booktitle={Proceedings of the 2018 on Asia Conference on Computer and Communications Security},
	pages={393--398},
	year={2018},
	doi={10.1145/3196494.3196504}
}

@article{guo2023searching,
	title={Searching Open-Source Vulnerability Function Based on Software Modularization},
	author={Guo, Xixi and Cai, Ruijie and Yin, Xiaokang and Shao, Wenqiang and Liu, Shengli},
	journal={Applied Sciences},
	volume={13},
	number={2},
	pages={701},
	year={2023},
	publisher={MDPI},
	doi={10.3390/app13020701}
}

@inproceedings{tang2020libdx,
	title={Libdx: A cross-platform and accurate system to detect third-party libraries in binary code},
	author={Tang, Wei and Luo, Ping and Fu, Jialiang and Zhang, Dan},
	booktitle={2020 IEEE 27th International Conference on Software Analysis, Evolution and Reengineering (SANER)},
	pages={104--115},
	year={2020},
	organization={IEEE},
	doi={10.1109/SANER48275.2020.9054845}
}

@inproceedings{backes2016reliable,
	title={Reliable third-party library detection in android and its security applications},
	author={Backes, Michael and Bugiel, Sven and Derr, Erik},
	booktitle={Proceedings of the 2016 ACM SIGSAC conference on computer and communications security},
	pages={356--367},
	year={2016},
	doi={10.1145/2976749.2978333}
}

@inproceedings{li2017libd,
	title={Libd: Scalable and precise third-party library detection in android markets},
	author={Li, Menghao and Wang, Wei and Wang, Pei and Wang, Shuai and Wu, Dinghao and Liu, Jian and Xue, Rui and Huo, Wei},
	booktitle={2017 IEEE/ACM 39th International Conference on Software Engineering (ICSE)},
	pages={335--346},
	year={2017},
	organization={IEEE},
	doi={10.1109/ICSE.2017.38}
}

@inproceedings{ma2016libradar,
	title={Libradar: Fast and accurate detection of third-party libraries in android apps},
	author={Ma, Ziang and Wang, Haoyu and Guo, Yao and Chen, Xiangqun},
	booktitle={Proceedings of the 38th international conference on software engineering companion},
	pages={653--656},
	year={2016},
	doi={10.1145/2889160.2889178}
}

@inproceedings{zhan2021atvhunter,
	title={Atvhunter: Reliable version detection of third-party libraries for vulnerability identification in android applications},
	author={Zhan, Xian and Fan, Lingling and Chen, Sen and We, Feng and Liu, Tianming and Luo, Xiapu and Liu, Yang},
	booktitle={2021 IEEE/ACM 43rd International Conference on Software Engineering (ICSE)},
	pages={1695--1707},
	year={2021},
	organization={IEEE},
	doi={10.1109/ICSE43902.2021.00150}
}

@inproceedings{zhan2020automated,
	title={Automated third-party library detection for android applications: Are we there yet?},
	author={Zhan, Xian and Fan, Lingling and Liu, Tianming and Chen, Sen and Li, Li and Wang, Haoyu and Xu, Yifei and Luo, Xiapu and Liu, Yang},
	booktitle={Proceedings of the 35th IEEE/ACM International Conference on Automated Software Engineering},
	pages={919--930},
	year={2020},
	doi={10.1145/3324884.3416582}
}

@inproceedings{zhang2019libid,
	title={Libid: reliable identification of obfuscated third-party android libraries},
	author={Zhang, Jiexin and Beresford, Alastair R and Kollmann, Stephan A},
	booktitle={Proceedings of the 28th ACM SIGSOFT International Symposium on Software Testing and Analysis},
	pages={55--65},
	year={2019},
	doi={10.1145/3293882.3330563}
}

@inproceedings{zhang2018detecting,
	title={Detecting third-party libraries in android applications with high precision and recall},
	author={Zhang, Yuan and Dai, Jiarun and Zhang, Xiaohan and Huang, Sirong and Yang, Zhemin and Yang, Min and Chen, Hao},
	booktitle={2018 IEEE 25th International Conference on Software Analysis, Evolution and Reengineering (SANER)},
	pages={141--152},
	year={2018},
	organization={IEEE},
	doi={10.1109/SANER.2018.8330204}
}

@article{newman2004fast,
	title={Fast algorithm for detecting community structure in networks},
	author={Newman, Mark EJ},
	journal={Physical Review E—Statistical, Nonlinear, and Soft Matter Physics},
	volume={69},
	number={6},
	pages={066133},
	year={2004},
	publisher={APS},
	doi={10.1103/PhysRevE.69.066133}
}

@inproceedings{asm2vec,
	title={Asm2vec: Boosting static representation robustness for binary clone search against code obfuscation and compiler optimization},
	author={Ding, Steven HH and Fung, Benjamin CM and Charland, Philippe},
	booktitle={2019 IEEE Symposium on Security and Privacy (SP)},
	pages={472--489},
	year={2019},
	organization={IEEE},
	doi={10.1109/SP.2019.00003}
}

@inproceedings{Gemini,
	title={Neural network-based graph embedding for cross-platform binary code similarity detection},
	author={Xu, Xiaojun and Liu, Chang and Feng, Qian and Yin, Heng and Song, Le and Song, Dawn},
	booktitle={Proceedings of the 2017 ACM SIGSAC Conference on Computer and Communications Security},
	pages={363--376},
	year={2017},
	doi={10.1145/3133956.3134018}
}

@inproceedings{ISRD,
	author       = {Xi Xu and
	Qinghua Zheng and
	Zheng Yan and
	Ming Fan and
	Ang Jia and
	Ting Liu},
	title        = {Interpretation-enabled Software Reuse Detection Based on a Multi-Level
	Birthmark Model},
	booktitle    = {43rd {IEEE/ACM} International Conference on Software Engineering,
	{ICSE} 2021, Madrid, Spain, 22-30 May 2021},
	pages        = {873--884},
	publisher    = {{IEEE}},
	year         = {2021},
	url          = {https://doi.org/10.1109/ICSE43902.2021.00084},
	doi          = {10.1109/ICSE43902.2021.00084},
	bibsource    = {dblp computer science bibliography, https://dblp.org}
}

@inproceedings{bingo,
	title={Bingo: Cross-architecture cross-os binary search},
	author={Chandramohan, Mahinthan and Xue, Yinxing and Xu, Zhengzi and Liu, Yang and Cho, Chia Yuan and Tan, Hee Beng Kuan},
	booktitle={Proceedings of the 2016 24th ACM SIGSOFT International Symposium on Foundations of Software Engineering},
	pages={678--689},
	year={2016},
	doi={10.1145/2950290.2950350}
}

@article{Binkit,
	title={Revisiting binary code similarity analysis using interpretable feature engineering and lessons learned},
	author={Kim, Dongkwan and Kim, Eunsoo and Cha, Sang Kil and Son, Sooel and Kim, Yongdae},
	journal={IEEE Transactions on Software Engineering},
	volume={49},
	number={4},
	pages={1661--1682},
	year={2023},
	publisher={IEEE},
	doi={10.1109/TSE.2022.3187689}
}

@misc{BinkitGithub,
	title = {{SoftSec-KAIST/BinKit: Binary Code Similarity Analysis Benchmark}},
	howpublished = "\url{https://github.com/SoftSec-KAIST/BinKit}",
	year = {2026}, 
	note = "[Online; accessed 9-August-2026]"
}

@misc{IDAPro,
	title = {{IDA Pro Disassembler and Debugger - Hex Rays}},
	howpublished = "\url{https://hex-rays.com/ida-pro}",
	year = {2026}, 
	note = "[Online; accessed 9-August-2026]"
}

@misc{readelf,
	title = {{readelf(1) — Linux manual page}},
	howpublished = "\url{https://man7.org/linux/man-pages/man1/readelf.1.html}",
	year = {2026}, 
	note = "[Online; accessed 9-August-2026]"
}

@misc{capstone,
	title = {{capstone on PyPI}},
	howpublished = "\url{https://pypi.org/project/capstone/}",
	year = {2026}, 
	note = "[Online; accessed 9-August-2026]"
}

@inproceedings{pang2021sok,
	title={Sok: All you ever wanted to know about x86/x64 binary disassembly but were afraid to ask},
	author={Pang, Chengbin and Yu, Ruotong and Chen, Yaohui and Koskinen, Eric and Portokalidis, Georgios and Mao, Bing and Xu, Jun},
	booktitle={2021 IEEE Symposium on Security and Privacy (SP)},
	pages={833--851},
	year={2021},
	organization={IEEE},
	doi={10.1109/SP40001.2021.00012}
}

@article{jia2022comparing,
	title={Comparing One with Many--Solving Binary2source Function Matching Under Function Inlining},
	author={Jia, Ang and Fan, Ming and Xu, Xi and Jin, Wuxia and Wang, Haijun and Tang, Qiyi and Nie, Sen and Wu, Shi and Liu, Ting},
	journal={arXiv preprint arXiv:2210.15159},
	year={2022},
	doi={10.48550/arXiv.2210.15159}
}

@article{jia20231,
	title={1-to-1 or 1-to-n? Investigating the Effect of Function Inlining on Binary Similarity Analysis},
	author={Jia, Ang and Fan, Ming and Jin, Wuxia and Xu, Xi and Zhou, Zhaohui and Tang, Qiyi and Nie, Sen and Wu, Shi and Liu, Ting},
	journal={ACM Transactions on Software Engineering and Methodology},
	volume={32},
	number={4},
	pages={1--26},
	year={2023},
	publisher={ACM New York, NY, USA},
	doi={10.1145/3561385}
}

@inproceedings{jia2024cross,
	title={Cross-inlining binary function similarity detection},
	author={Jia, Ang and Fan, Ming and Xu, Xi and Jin, Wuxia and Wang, Haijun and Liu, Ting},
	booktitle={Proceedings of the IEEE/ACM 46th International Conference on Software Engineering},
	pages={1--13},
	year={2024},
	doi={10.1145/3597503.3639080}
}

@inproceedings{qiu2015library,
	title={Library functions identification in binary code by using graph isomorphism testings},
	author={Qiu, Jing and Su, Xiaohong and Ma, Peijun},
	booktitle={2015 ieee 22nd international conference on software analysis, evolution, and reengineering (saner)},
	pages={261--270},
	year={2015},
	organization={IEEE},
	doi={10.1109/SANER.2015.7081836}
}

@article{qiu2015using,
	title={Using reduced execution flow graph to identify library functions in binary code},
	author={Qiu, Jing and Su, Xiaohong and Ma, Peijun},
	journal={IEEE Transactions on Software Engineering},
	volume={42},
	number={2},
	pages={187--202},
	year={2016},
	publisher={IEEE},
	doi={10.1109/TSE.2015.2470241}
}

@article{ahmed2021learning,
	title={Learning to find usages of library functions in optimized binaries},
	author={Ahmed, Toufique and Devanbu, Premkumar and Sawant, Anand Ashok},
	journal={IEEE Transactions on Software Engineering},
	volume={48},
	number={10},
	pages={3862--3876},
	year={2022},
	publisher={IEEE},
	doi={10.1109/TSE.2021.3106572}
}

@inproceedings{lin2024reifunc,
	title={ReIFunc: Identifying Recurring Inline Functions in Binary Code},
	author={Lin, Wei and Guo, Qingli and Yu, DongSong and Yin, Jiawei and Gong, Qi and Gong, Xiaorui},
	booktitle={2024 IEEE International Conference on Software Analysis, Evolution and Reengineering (SANER)},
	pages={670--680},
	year={2024},
	organization={IEEE},
	doi={10.1109/SANER60148.2024.00074}
}

@article{sha2025optrans,
	title={OpTrans: enhancing binary code similarity detection with function inlining re-optimization},
	author={Sha, Zihan and Lan, Yang and Zhang, Chao and Wang, Hao and Gao, Zeyu and Zhang, Bolun and Shu, Hui},
	journal={Empirical Software Engineering},
	volume={30},
	number={2},
	pages={49},
	year={2025},
	publisher={Springer},
	doi={10.1007/s10664-024-10605-x}
}

@article{binosi2023bino,
	title={Bino: Automatic recognition of inline binary functions from template classes},
	author={Binosi, Lorenzo and Polino, Mario and Carminati, Michele and Zanero, Stefano},
	journal={Computers \& Security},
	volume={132},
	pages={103312},
	year={2023},
	publisher={Elsevier},
	doi={10.1016/j.cose.2023.103312}
}

@inproceedings{lin2023fsmell,
	title={FSmell: recognizing inline function in binary code},
	author={Lin, Wei and Guo, Qingli and Yin, Jiawei and Zuo, Xiangyu and Wang, Rongqing and Gong, Xiaorui},
	booktitle={European Symposium on Research in Computer Security},
	pages={487--506},
	year={2023},
	organization={Springer},
	doi={10.1007/978-3-031-51476-0_24}
}

@article{dall2022highliner,
	author = {Dall'Aglio, Lorenzo and Binosi, Lorenzo and Carminati, Michele and Zanero, Stefano and Polino, Mario},
	title = {Highliner: Enhancing Binary Analysis through NLP-Based Instruction-Level Detection of C++ Inline Functions},
	year = {2025},
	issue_date = {November 2025},
	publisher = {Association for Computing Machinery},
	address = {New York, NY, USA},
	volume = {28},
	number = {4},
	issn = {2471-2566},
	url = {https://doi.org/10.1145/3765521},
	doi = {10.1145/3765521},
	journal = {ACM Trans. Priv. Secur.},
	month = oct,
	articleno = {51},
	numpages = {22}
}

@inproceedings{10.1145/3533767.3534367,
	author = {Wang, Hao and Qu, Wenjie and Katz, Gilad and Zhu, Wenyu and Gao, Zeyu and Qiu, Han and Zhuge, Jianwei and Zhang, Chao},
	title = {jTrans: jump-aware transformer for binary code similarity detection},
	year = {2022},
	isbn = {9781450393799},
	publisher = {Association for Computing Machinery},
	address = {New York, NY, USA},
	url = {https://doi.org/10.1145/3533767.3534367},
	doi = {10.1145/3533767.3534367},
	booktitle = {Proceedings of the 31st ACM SIGSOFT International Symposium on Software Testing and Analysis},
	pages = {1–13},
	numpages = {13},
	location = {Virtual, South Korea},
	series = {ISSTA 2022}
}

@misc{GCC_inline,
	title = {{Optimize Options (Using the GNU Compiler Collection (GCC))}},
	howpublished = "\url{https://gcc.gnu.org/onlinedocs/gcc/Optimize-Options.html}",
	year = {2026}, 
	note = "[Online; accessed 9-August-2026]"
}

@inproceedings{abusabha2025deep,
	title={A Deep Dive into Function Inlining and its Security Implications for ML-based Binary Analysis},
	author={Abusabha, Omar and Uhm, Jiyong and Abuhmed, Tamer and Koo, Hyungjoon},
	booktitle={Proceedings of the 2026 Network and Distributed System Security Symposium (NDSS)},
	year={2026},
	doi={10.14722/ndss.2026.241872}
}

@article{jia2025towards,
	title={Towards an Oracle for Binary Decomposition Under Compilation Variance},
	author={Jia, Ang and Jiang, He and Ren, Zhilei and Li, Xiaochen and Yang, Zhipeng and Duan, Yaxin and Fan, Ming and Liu, Ting},
	journal={ACM Transactions on Software Engineering and Methodology},
	year={2026},
	publisher={ACM},
	doi={10.1145/3803807}
}

\end{document}